\documentclass[12pt]{iopart}

\usepackage{iopams}

\usepackage{blindtext}
\usepackage{graphicx}
\usepackage{verbatim}
\usepackage{color}
\usepackage{microtype}
\usepackage{placeins}
\usepackage{bm}
\usepackage{amsfonts}
\usepackage{amsbsy}
\usepackage{tabularx}
\usepackage{floatrow}   
\usepackage[normalem]{ulem}
\usepackage{tikz}

\newcommand{\beq}{\begin{equation}}
\newcommand{\eeq}{\end{equation}}
\newcommand{\bqa}{\begin{eqnarray}}
\newcommand{\eqa}{\end{eqnarray}}
\newcommand{\bal}{\begin{equation}\begin{aligned}}
\newcommand{\eal}{\end{aligned}\end{equation}}

\newcommand{\erf}[1]{(\ref{#1})}

\newcommand{\erfs}[2]{(\ref{#1})--(\ref{#2})}

\newcommand{\srf}[1]{section~\ref{#1}}

\newcommand{\frf}[1]{figure~\ref{#1}}

\newcommand{\dg}{^\dagger}

\newcommand{\ket}[1]{\left|#1\right\rangle}

\newcommand{\half}{\frac{1}{2}}

\newcommand{\blk}{\color{black}}

\newcommand\stPW{\bgroup\markoverwith{\textcolor{red}{\rule[0.5ex]{2pt}{0.4pt}}}\ULon}

\begin{document}
\title[Tomography of an optomechanical oscillator via parametric amplification] {Tomography of an optomechanical oscillator via parametrically amplified position measurement}

\date{\today}

\author{P.~Warszawski\textsuperscript{1}, A.~Szorkovszky\textsuperscript{2,3}, W.~P.~Bowen\textsuperscript{2} and A.~C.~Doherty\textsuperscript{1}}

\address{\textsuperscript{1} Centre for Engineered Quantum Systems, School of Physics,
The University of Sydney, Sydney, NSW 2006, Australia}
\address{\textsuperscript{2} Centre for Engineered Quantum Systems, School of Mathematics and Physics,
The University of Queensland, Brisbane, Queensland 4072, Australia}
\address{\textsuperscript{3} Institute Mittag-Leffler, The Royal Swedish Academy of Sciences, S-18262 Djursholm, Sweden}

\begin{abstract}

We propose a protocol \blk for quantum state tomography of nonclassical states in optomechanical systems. Using a parametric drive, the procedure overcomes the challenges of weak optomechanical coupling, poor detection efficiency, \blk and thermal noise to enable \blk high efficiency homodyne measurement.  Our analysis is based on the analytic description of the generalized measurement that is performed when optomechanical position measurement competes with thermal noise and a parametric drive. The proposed experimental procedure is numerically simulated in realistic parameter regimes, which allows us to show that tomographic reconstruction of otherwise unverifiable nonclassical states is made possible.

\end{abstract}


\maketitle
\section{Introduction} 
\label{sec:introduction}

Optomechanical systems, consisting of interacting electromagnetic and mechanical modes, make accessible the preparation and control of macroscopic nonclassical states~\cite{revCavOptMech,teufel2011sideband,schliesser2008resolved,o2010quantum,PhysRevA.82.061804}.  Squeezed states~\cite{PhysRevA.79.063819,purdy2013strong,wollman2015quantum} can be created using a linearized optomechanical coupling, while using non-linear measurement, or a non-linear coupling, allows the synthesis of more exotic non-Gaussian states~\cite{PhysRevLett.107.020405,ludwig2012enhanced}.  Nonclassical optomechanical states are of interest for many reasons; as a macroscopic test of quantum mechanics, as a resource for quantum metrology~\cite{Schreppler1486} and quantum information processing~\cite{schmidt2012optomechanical}, and as a way to investigate hybrid quantum systems~\cite{1402.1195}.  Undoubtedly, the creation of such states is a major experimental challenge.  However, in this paper we focus on the complimentary task of verifying that the claimed state has been produced.

The gold standard for demonstrating that a nonclassical state has been prepared is quantum state tomography~\cite{tomRevLvovsky,tomRevAriano}.  This process is performed by measuring many copies of an identically prepared state of interest.  The measurement must be varied across the repetitions such that a quorum of (non-commuting) observables is probed~\cite{fano}.  For example, in optical homodyne tomography the quorum is formed by varying the local oscillator phase and thus measuring the different field quadratures.  In contrast, the related technique of heterodyne detection involves simultaneously jointly measuring two quadratures at the expense of additional noise.  Repeated measurements on identically prepared states allows a statistical picture to be developed, with this information captured in the density matrix assigned to the prepared state.  The number of copies necessary to achieve a particular level of certainty regarding the prepared state depends on the experimental procedure comprising the measurement in combination with the nature of the state itself.

In the setting of optomechanics, tomography can be performed on the basis of continuous measurement of the position of the oscillator. Although, at any given moment, such a measurement reads out only the position of the oscillator, by monitoring the position over time it is possible to estimate all quadrature observables of the initial state and thereby obtain the required quorum of observables. There is significant prior work on using continuous measurements and time dynamics in this way to perform quantum tomography, particularly the early work by Deutsch and Jessen and collaborators which introduced the idea of continuous measurement tomography in the context of reconstructing the hyperfine state of an ensemble of atoms theoretically~\cite{silberfarb2005quantum,riofrio2011quantum,cook2014single} and implemented it in practice~\cite{smith2004continuous,smith2013quantum}.  
To model our measurements, we will apply the standard tools of the trajectory theory of continuous quantum measurements~\cite{WisMil10} and use it to determine the initial quantum state at the beginning of a measurement. Continuous measurement tomography where the system dynamics are used to achieve a quorum of observables for quantum state tomography have also been considered in~\cite{six2016quantum,chantasri2018quantum}

Our approach to this will be to explicitly determine the {\it positive operator valued measure} (POVM) that is implemented by the continuous optomechanical position measurement, following~\cite{jacLin,wisQTraj,herkommer1996localization}. This general approach is very similar to that of~\cite{six2016quantum,chantasri2018quantum} where continuous measurement state tomography of qubits is considered. There are several other approaches to obtaining information about initial states in continuous quantum measurement that are closely related, for example~\cite{gammelmark2013past,guevara2015quantum,tan2015prediction,hacohen2016quantum}. Recent work also demonstrates optimal tomography of qubit states by sequential weak measurements on an ensemble by demonstrating that it is possible to implement the spin-coherent-state in this way~\cite{shojaee2018optimal}.


In optomechanical systems, measurement of the mechanical mode is achieved by detecting the light leaking out of the coupled optical mode, which is phase shifted by an amount that depends on the mechanical position.  
In the typical regime, the coupling is not large enough to resolve the mechanical oscillator on time scales short compared with the period of motion. 
 Consequently, in this regime, the continuous measurement of position monitors both quadratures of the mechanical motion equally, such that the measurement is similar to optical heterodyne detection~\cite{sme}. In addition to the noise penalty associated with heterodyne detection, the signal-to-noise ratio is also degraded by thermal fluctuations that enter from the mechanical bath during the measurement period as well as quantum backaction noise due to the non-perfect quantum efficiency of the optical or microwave detector.
This poses significant challenges for establishing the density matrix of non-classical states of motion. One approach that has been developed to address these challenges is to use a stroboscopic interaction to achieve a backaction evading measurement~\cite{Suh1253258,Vasilakis2015}. 
So long as the interaction is strong enough that the measurement rate dominates the rate at which thermal fluctuations enter the oscillator, and the detection efficiency is sufficiently high, such a measurement can approach the equivalent of ideal optical homodyne detection~\cite{bowMil}. Another suggested approach has been to imprint the mechanical state onto a strong optical coherent pulse~\cite{Shahandeh}, although their analysis neglects the thermal fluctuations.

In this paper, we consider an alternative approach that retains the usual continuous optical measurement, but includes 
an additional resonant parametric amplification of the mechanical oscillator.
Parametric amplification plays an important role in the manipulation and measurement of oscillator quantum states, by squeezing one quadrature and amplifying the other.  It is a readily available resource in optomechanics, since mechanical systems often possess significant nonlinearities that can be used to enable parametric driving of the oscillator~\cite{szorkovszky2014detuned,unterreithmeier2009universal,suh2010parametric,castellanos2008amplification,
rugar1991mechanical}.  Furthermore, in~\cite{szorkovszky2012position,PhysRevLett.107.213603,szorkovszky2013strong} it is shown that, in conjunction with continuous measurement, arbitrarily high levels of squeezing can be achieved in principle. This proposal is closely analogous to a recent experiment in circuit QED has used parametric drive of a microwave cavity to improve the effective efficiency of measurements on transmon qubits~\cite{eddins2018high}.

Importantly for quantum state tomography, parametric amplification adds no noise, in principle, and is therefore very effective at transducing small signals. Indeed, its use to improve tomography of optical field states was proposed more than two decades ago~\cite{leonhardt1994high}. Note, also, that recently a state of the art experiment based on the ideas of~\cite{leonhardt1994high} has been proposed~\cite{1367-2630-20-1-013005}.
When applied to tomography of mechanical oscillator states, however, significant differences are readily apparent. 
The oscillator is a dynamical system that can be probed repeatedly. 
As such, a series of measurements, or a continuous measurement, can be made. How can these multiple measurement outcomes be best used to estimate the initial state of the oscillator?
How is the information gathered from one measurement influenced by the quantum backaction due to measurements in the past? Finally, if the mechanical oscillator is connected, as usual, to a thermal bath, how does this effect the optimal approach to obtain estimates, and their accuracy? These are the main questions addressed in this paper.

In this paper we make use of two complimentary methods of analysis in order to obtain simple analytical expressions for the generalized measurement, or  POVM, associated with a mechanical oscillator exposed to different sequences of continuous measurement and parametric driving.  The first, simplified, method models the measurement part of the dynamics as being completely analogous to heterodyne detection of some specific, but unknown efficiency, for which the POVM is well known.  Important conclusions can then be drawn regarding the effectiveness of the measurement protocols that involve parametric amplification.  However, this initial analysis is non-constructive, in the sense that it does not specify the effective heterodyne efficiency that is achieved in the optomechanical experiment, or the appropriate filter of the measured current required to achieve an optimal estimate of the initial system state.  To obtain this we use, as discussed above, the theory of continuous quantum measurements, and, specifically, stochastic master equations or quantum trajectories~\cite{WisMil10,sme}.  These equations specify the stochastic evolution of the mechanical quantum state given the measurement record.  In addition to this theoretical framework, we make use of our recent analysis~\cite{warDoh} in order to find the more detailed form of the POVM.
Our results show, broadly, that
a sufficiently strong parametric drive converts  a continuous optomechanical measurement into near-perfect homodyne detection, with the detected quadrature determined by the phase of the parametric drive. We are able to find explicit expressions for the strength of the parametric drive that is required to approach ideal homodyne measurement in terms of system parameters such as the quantum efficiency of the optical measurement, the damping rate of the oscillator and the thermal phonon number. Importantly, neither high detector efficiency nor high optomechanical couplings are required to reach this regime. All that is required is that the parametric drive can be made strong enough to outpace the effects of the thermal bath on the mechanical oscillator. We show that this result holds in two regimes of optomechanics: the bad-cavity regime with on-resonant drive, and the sideband-resolved regime with blue-detuned drive. Although details differ, the physics is very similar in these two experimentally relevant regimes. This opens a pathway to efficiently perform quantum state tomography of mechanical systems in  new parameter regimes, where strong measurements and high detector efficiency are unavailable.

Although increasing the resonant parametric amplification strength transforms the POVM from a heterodyne to homodyne form, it is not a priori obvious that this actually leads to superior tomographic performance.  The intuition for it to do so is that, for some states that we wish to reconstruct, it will be more effective, in the probing of each state copy, to gain precise information regarding a particular quadrature than imprecise information concerning both.  This is expected to be manifest for states that have hard to resolve features, or, equivalently, quickly oscillating Wigner function representations.  The prototypical example of such states are the ``cat state" superpositions, so we choose to base our numerical investigation around these.  The simulation of experimental data allows a comparison of the tomographic performance as a function of the strength of the resonant parametric amplification.  Our findings confirm the expectation that performance is greatest when information is concentrated in a single quadrature, with the homodyne paradigm being most desirable.

The organisation of this paper is as follows.  In \srf{smeSoln} the system is introduced in detail and the mathematical techniques necessary to prepare for a direct analysis of tomographic performance are described.  Most of the general conclusions are drawn in \srf{simple}, where a simplified analysis of the measurement protocol is performed.  The limitations of this analysis are then addressed in \srf{tomogram} and \srf{BlueDetuned}, where specific expressions are provided for two experimental regimes of interest.
Then, numerical simulations are studied in \srf{sims} and, finally, conclusions are provided in \srf{conc}.

\section{Model of optomechanical position measurement}
\label{smeSoln}
We consider a conventional optomechanical position measurement~\cite{revCavOptMech}. Our results apply equally to nanoelectromechanical systems with detection of a microwave field but, to be concrete, we will refer to optical fields throughout. The system is indicated schematically in \frf{schematic}(a). A laser of angular frequency $\omega_{{\rm L}}$ drives a single mode of an optical cavity having angular frequency $\omega_{{\rm cav}}$. The laser-cavity detuning is $\Delta=\omega_{{\rm L}}-\omega_{{\rm cav}}$.  
The mechanical oscillator ($\omega_{{\rm m}}$) induces a shift in the resonant frequency of the optical cavity that is proportional to its position. The optical damping rate is $\kappa$, while the mechanical mode is taken to be coupled to a finite temperature thermal bath described by damping rate $\gamma$ and thermal phonon occupancy $n_{{\rm th}}$.  The light-enhanced optomechanical coupling strength, $g$, originates from the radiation pressure of cavity photons on the mechanical oscillator. It is defined as $g=\sqrt{N} G x_{\rm zp}$ where $N$ is the mean steady-state intracavity photon number, $G$ is the the optical frequency shift per unit displacement of the mechanical oscillator and $x_{\rm zp}$ is root-mean-square displacement of the ground state wavefunction of the oscillator, i.e. its zero-point motion amplitude. \blk
 We restrict our attention to the regime where the rotating-wave approximation (RWA) for the optics and the `linearization' approximation for the optomechanical coupling can be made~\cite{revCavOptMech}. We will work in units such that $\hbar=1$.

In order to facilitate an analytic treatment of the POVMs that arise as a result of the continuous measurement, we will study regimes of the optomechanical system where the cavity field relaxes rapidly compared to the mechanical dynamics, i.e. $\kappa \gg \gamma$, where $\kappa$ and $\gamma$ are the respective decay rates of the optical cavity and mechanical oscillator. This is the regime in which the vast majority of opto- and electromechanics experiments reside, an exception being recent work where the mechanical oscillator is coupled to a second cavity  to greatly increase its decay rate~\cite{Toth2017}.
Consequently, the system can be approximately described by a master equation involving the mechanics alone. (We expect our protocol to work qualitatively similarly in other cases but defer a study of this to future work.) 
We consider two specific experimentally relevant regimes \blk in this paper: the zero-detuned bad-cavity regime ($\Delta=0$ and $\kappa\gg \omega_{{\rm m}}$ \blk) and the blue-detuned resolved-sideband regime \blk ($\Delta\approx \omega_{{\rm m}}$ and $\omega_{{\rm m}}\gg\kappa$). In both regimes we 
require weak coupling $\kappa\gg g$, such that \blk  the dynamics of the mechanical mode alone can be considered by adiabatically eliminating the cavity. 
In this limit, the optomechanical coupling is well-described by the measurement rate $\mu=4g^2/\kappa$ \cite{bowMil}.
$\mu$ defines an the upper bound to the rate at which information about the mechanical system can be extracted from the optical field leaving the cavity. In the zero-detuned bad-cavity regime, if the measurement rate exceeds the thermal decoherence rate $\gamma n_{\rm th}$ (or expressed in terms of the {\it optomechanical cooperativity} $C=g^2/\kappa\gamma$,  $C>n_{\rm th}$), the quantum back-action heating associated with the measurement exceeds the heating from the mechanical thermal bath, as first observed in~\cite{zdexperiment}. \blk
In the blue-detuned regime, laser light entering the cavity is accompanied by the absorption of a phonon by the oscillator, leading to amplification. See~\cite{BDexpI,BDexpII,bdexperiment} for some blue-detuned experiments that make use of this. We will be able to compare the effect of this intrinsic amplification with the effect of the parametric drive on the tomographic signal.

%

The general experimental procedure, illustrated in \frf{schematic}(a), is to shine laser light upon the optical cavity, which then leads to a phase shift on the reflected light that depends on the mechanical oscillator position. Monitoring the returning light through homodyne detection results in a measured current, $I(t)$, allowing information about the mechanical oscillator position to be gathered.  Our work is concerned with tomography, so only the initial mechanical oscillator state is of interest, but despite this the measurement takes place over a finite amount of time corresponding to, in the regimes of interest in this paper, many mechanical oscillation periods.  The presence of the thermal bath adds noise to the system, as does the radiation pressure fluctuations that represent the backaction noise of the measurement, so that typically it is only at early times that the monitored light will contain a significant amount of tomographic information.  Statistics are collected over a large number of trials that are then used to perform quantum state tomography.

\begin{figure}
	\centering
\includegraphics[width=1\hsize,scale=1]{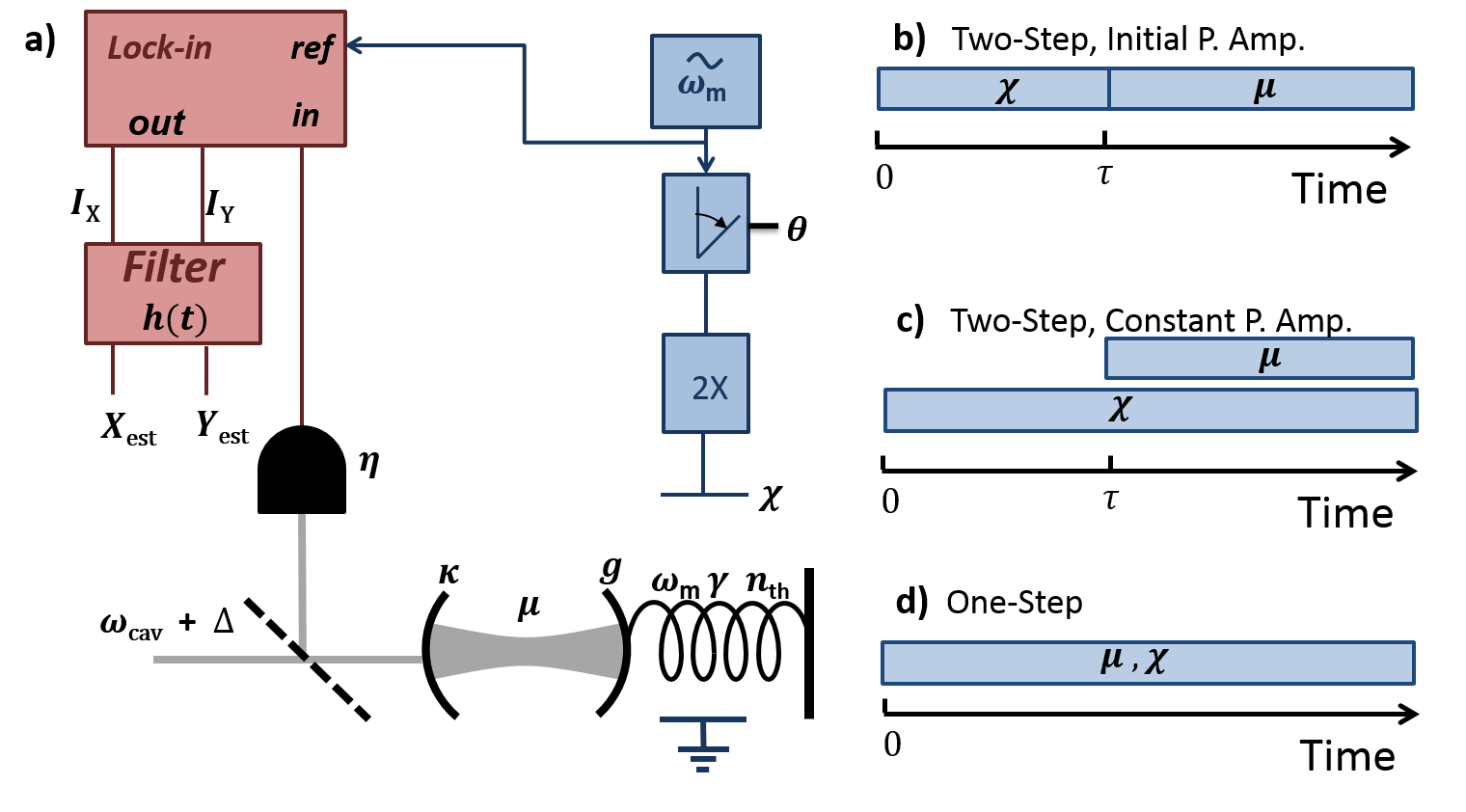}
	\caption{\label{angles} {\bf a)} Schematic of a driven optomechanical system subject to continuous weak measurement.  The measured signal $I(t)$ is demodulated into currents $I_X, I_Y$ representing the $X$ and $Y$ quadratures.  The currents can be filtered, by $h(t)$, to provide an estimate $X_{{\rm est}},Y_{{\rm est}}$ of the initial quadratures of the quantum state that is to be reconstructed using tomography.  The spring constant of the mechanical oscillator is capacitively modulated at strength $\chi$ at twice the oscillator frequency, $\omega_m$, which gives resonant parametric amplification and leads to system squeezing of a quadrature defined by the modulation phase $\theta$.  The oscillator is coupled to a finite temperature ($n_{{\rm th}}$) bath at rate $\gamma$.  It is also coupled, through optomechanical coupling, $g$, to an optical cavity, of frequency $\omega_{{\rm cav}}$, which has a damping rate of $\kappa$.  A laser, of frequency $\omega_{{\rm L}}=\omega_{{\rm cav}}+\Delta$, is shone onto the cavity, with the reflected light directed into the detector.  
{\bf b-d)}~In these schematics we show how the one and two-step experiments are differentiated in terms of the timing of measurement and parametric amplification (abbreviated as P. Amp.).  The two-step process switches on the measurement, $\mu$, after an amplification time $\tau$.  At time $\tau$ in the two-step process we consider both maintaining the squeezing for the duration of the measurement (subplot c) and turning it off (subplot b).  
The one-step process (subplot d) has both constant measurement, $\mu$, and constant amplification, $\chi$.  
}
\label{schematic}
\end{figure}


To combat the degradation of the signal due to the thermal noise, and allow a superior signal to noise ratio to be achieved, we propose to selectively amplify a chosen quadrature of the quantum state.
The schematic in \frf{schematic}(a) shows the presence of this additional experimental resource: the ability to resonantly parametrically drive the oscillator. We use the quadrature definitions $X=(a+a\dg)/\sqrt{2}$ and $Y=-i(a-a\dg)/\sqrt{2}$ in terms of the interaction picture annihilation and creation operators of the oscillator. The parametric drive allows a quadrature of the mechanical motion to be squeezed with strength $\chi$.  It is modelled by the squeezing Hamiltonian
\beq
H=\frac{i\chi}{4}\left(e^{-i \theta}a^2-e^{i \theta}a^{\dagger 2}\right),
\label{hamiltonian}
\eeq
with the angle $\theta$ determining which quadrature is squeezed ($\theta=0$ squeezes the $X$ quadrature, $\theta=\pi$ squeezes the $Y$ quadrature).  Experimentally, the squeezing can be achieved, for instance, by electromechanically modulating the frequency of the oscillator at twice its resonant frequency, $\omega_m$, with a phase of $\theta$ relative to that of the laser~\cite{PhysRevLett.107.213603,rugar1991mechanical}.  Prior to the application of the squeezing, and within the rotating wave approximation that we employ, there is no preferred mechanical quadrature and an equal amount of information is gathered about each in any single trial.  The squeezing removes this symmetry and allows a closer inspection of one quadrature at the cost of another.  It is the amplified, or {\it anti}-squeezed, quadrature that will be more visible in the measurement record and thus contain the bulk of the tomographic information.  

The experimentalist has access to the complete measurement record $I(t)$. We can think of this measurement record as arising from a sequence of very weak measurements of the system occurring at each moment of time. We will often refer to this sequence of measurements as a composite measurement. To find the POVM that represents a composite measurement, the theory of open quantum systems and quantum trajectories can be applied~\cite{WisMil10,wisQTraj,jacSteck}.  Information about the mechanical system leaks out into the environment, which is both detected (via the optomechanical measurement scheme) and undetected (information leaking into the thermal bath or lost due to optical measurement inefficiencies\blk). This requires a density matrix description in which the state matrix follows a different trajectory in each run of the experiment. This makes it more challenging to determine the POVM that arises from the composite measurement, relative to well-known examples in the literature~\cite{wisQTraj,jacSteck}.  Recently, two of the authors have given a general method of solution for Gaussian bosonic quantum trajectories such as the ones of interest here~\cite{warDoh}. This makes it possible to determine the relevant POVMs analytically in the two regimes of interest. We detail our methods further in \ref{qtraj}.  

While the statistics of the composite measurement can, in principle, depend on any property of the measurement record $I(t)$, we find, as in earlier examples~\cite{wisQTraj,jacSteck}, that there are two simple sufficient statistics that essentially determine the POVM. These correspond to each quadrature of the oscillator motion, and the general form of the sufficient statistic for the $X$-quadrature is
\beq
X_{{\rm est}}=\int^t_0 h_X(r)I(r)dr,
\label{filterGeneral}
\eeq
where $h_{X}$ is a filter function that depends on the system parameters. There is a corresponding form for $Y_{\rm est}$ with a possibly distinct filter function $h_{Y}$.  The quantities $X_{\rm est},Y_{\rm est}$ are estimates of the {\it initial} system quadratures as we wish to perform tomography.  The integral receives dominant contributions from early times as the system state will progressively become less correlated with the initial state due to measurement back-action and coupling to decohering baths.  Finding the form of $h_X(t)$, for the non-trivial system of interest, is an important result of our work and specific expressions will be given in later sections.  The origin of the term `composite measurement' is made clear from the form of \erf{filterGeneral}, where the measurement results for each small time slice are integrated over the duration of the measurement.

It is easiest to specify the POVM, that is implemented by the composite optomechanical position measurement, by considering an initial coherent state. We denote the POVM by $W_I$ and the complex amplitude by $\alpha_0$. 
From the resultant Gaussian statistics, it will be made clear where the composite measurement lies in relation to homodyne and heterodyne detection. Moreover, the statistics of the POVM for an initial coherent state completely determine the POVM for all initial states, so this is without loss of generality. (To see this recall that a density matrix $\rho$ is completely specified by its Q-function $Q(\alpha)=\langle \alpha |\rho |\alpha\rangle$.) The results of~\cite{warDoh} imply a bivariate Gaussian probability distribution for $X_{\rm est}$ and $Y_{\rm est}$: 
\beq
  \fl P(X_{\rm est},Y_{\rm est}|\alpha_0)=\langle \alpha_0 | W_{I}|\alpha_0\rangle =
N \exp\left[-\frac{1}{2(1-\rho_{c}^{2})}\left(\frac{\Delta_{X}^2}{\sigma_{X}^{2}}+\frac{\Delta_{Y}^2}{\sigma_{Y}^{2}}-\frac{2\rho_{c}\Delta_{X}\Delta_{Y}}{\sigma_{X}\sigma_{Y}}\right) \right],
\label{eq:POVM}
\eeq
where we have broken $\alpha_0$ into quadratures as $\alpha_0=\left(X_0+iY_0\right)/\sqrt{2}$ and defined $\Delta_{X}=X_{\rm est}-X_0$ and $\Delta_{Y}=Y_{\rm est}-Y_0$.
$\sigma_{X}^2$ and $\sigma_{Y}^2$ are the observed variances of $X_{\rm est}$ and $Y_{\rm est}$ respectively.  The bivariate correlation, $\rho_{c}$, is equal to zero if the squeezing is aligned with the $X$ or $Y$ quadratures. $\sigma_{X}$, $\sigma_{Y}$ and  $\rho_{c}$ are determined by the underlying system parameters and we defer explicit expressions until later sections. $N$ is a normalization factor that does not depend on $X_0$ or $Y_0$. With our chosen quadrature operator definitions, 
 the POVM of (\ref{eq:POVM}) corresponds to ideal heterodyne detection when $\sigma_X^2=1=\sigma_Y^2$ and $\rho_c=0$. Ideal homodyne detection of the $Y$-quadrature corresponds to $\sigma_Y^2=1/2$ while $\sigma_X^2=\infty$ and $\rho_c=0$. Imperfect heterodyne detection with overall efficiency $\eta$ corresponds to $\sigma_X^2=1/\eta=\sigma^2_Y$, while homodyne detection of $Y$ with efficiency $\eta$ corresponds to $\sigma^2_Y=1/2\eta$. 
It should be observed that for ideal homodyne detection $\sigma_Y^2$ exactly matches the coherent state variance, introducing no uncertainty over-and-above the inherent uncertainty of the Y-quadrature. On the other hand, heterodyne detection results in one unit of zero-point uncertainty above the intrinsic quadrature uncertainty. This is, of course, a direct result of the Heisenberg uncertainty principle which forbids perfect simultaneous measurements of non-commuting observables.  \blk

We can now be more concrete about the content of the following sections.  Firstly, we obtain as much as possible of the important physics via a simplified analysis that treats the measurement step of the protocol as being equivalent to an instantaneous inefficient heterodyne measurement.  Then, to delve deeper into the impact of a more realistic model of our proposed measurement protocol, and also find the form of the filters in \erf{filterGeneral}, we give an explicit analysis in 
two different regimes: in section~\ref{tomogram} the zero-detuning bad-cavity regime is analysed in some detail, and then, in section~\ref{BlueDetuned}, a similar, but briefer, analysis is performed for the blue-detuned regime. Next, we conduct an explicit analysis of quantum state tomography for non-classical mechanical states using the proposed multi-stage measurement protocol.  The tomographic data analysis procedure is well understood for heterodyne and homodyne measurement, the results of which can be readily converted into an estimate of the initial mechanical oscillator state~\cite{tomRevAriano}.  The situation of inefficient detection, and possibly additional noise directly added to the output signal, can also be described within this framework.  Of course, with tomography, the input state is not known, so the problem is to invert from the measured probabilities back to the unknown state with the aid of the POVM.  Our preferred method of doing so is via Maximum Likelihood (MaxLik) tomography~\cite{tomRevLvovsky,ORIGINALmaxLik}. We will implement this measurement on simulated  experimental data in \srf{sims}. 
\blk

\section{Simplified analysis of optomechanical measurement with parametric amplification}
\label{simple}

In this section, we provide a simplified analysis of optomechanical tomographic performance in the case of a two-step protocol (see \frf{schematic}(b)) that is followed in order to extract information concerning the initial state of the mechanical oscillator.  The two steps are, firstly, a period of parametric amplification and, secondly, the optomechanical measurement of the oscillator, via the detection of light leaking from the cavity, as per \frf{schematic}(a).  The simplification that we employ is to model the measurement step in a broad sense, so that although the expressions we provide are accurate, they do not display an explicit dependence upon the intra-measurement dynamics.  In later sections, these dependences will be revealed.  The purpose for the current simplification is that important physics relating to the impact of the supplementary squeezing step can still be introduced.  We will highlight and discuss the transformation of the measurement protocol from having a purely heterodyne description, with the squeezing turned off, to having strong homodyne features, when the squeezing is on.  We wish to be completely clear in our use of the terms `measurement protocol', which can include  supplementary experimentally implemented processes in addition to physical measurement --- that is, an initial squeezing step is included --- whereas `measurement step' refers to the well defined portion of the protocol during which the measurement strength is non-zero.

%
%
%

To begin, we familiarize ourselves with the description of the second step (the measurement step).  As has been discussed, the optomechanical measurement in the {\it absence} of squeezing (step-two of \frf{schematic}(b)) provides an equal amount of information about both mechanical quadratures. This is true in both the zero-detuned and blue-detuned regimes. However, the measurement's quality is degraded by the prolonged exposure of the system to the thermal bath.  Additionally, the efficiency $\eta$ of the detector must be considered, and one can also envision the measurement being carried out for some finite (and therefore sub-optimal) time $T$.  It is therefore natural to describe the measurement step --- step-two of the protocol --- in terms of non-ideal heterodyne detection, having some effective efficiency $\eta_{\rm het}$, which describes the overall noise level of the measurement.  To be clear, all the measurement imperfections are included in $\eta_{\rm het}$ (a function of $\gamma, n_{\rm th}, \mu,\eta,T$) and it will be, in general, smaller than the actual detector efficiency, $\eta_{\rm het}\leq\eta$.  In later sections, explicit expressions for $\eta_{\rm het}$ will be provided, for both the zero-detuned and blue-detuned regimes, but in this section we will focus on general results independent of the detailed specification of the measurement step.  

To provide the reader more clarity regarding the role of $\eta_{\rm het}$, let us analyse a heterodyne measurement a little further.
The outcome of a heterodyne measurement provides an unbiased estimator of the means of system quadratures but with an, inevitably, increased variance due to the joint measurement of non-commuting variables. The amount of excess noise for perfect heterodyne detection is 1/2, while for an inefficient detector it is given by $\frac{2- \eta_{{\rm het}}}{2\eta_{{\rm het}}}$~\cite{tomRevAriano}.  Thus, the heterodyne measurement yields two Gaussian random variables, $x,y$, with $y$ having variance $V_{Y}+\frac{2- \eta_{{\rm het}}}{2\eta_{{\rm het}}}$ (similarly for $x$), with $V_{Y}$ being the variance of the $Y$ quadrature at the time of the measurement.  One can see that the measurement result variance is the sum of two components: an initial variance and excess noise.  It is this excess noise that makes a heterodyne estimate noisier than a homodyne estimate. 
The statistics of the heterodyne measurement of a coherent state are given in \erf{eq:POVM}, with the specification that $\sigma_{{\rm X}}^{2}=\sigma_{{\rm Y}}^{2}=1/\eta_{\rm het}$ and $\rho_{c}=0$, however, we emphasize that the POVM associated with a measurement is in one-to-one correspondence with these measurement statistics so that both representations are  equivalent and we refer to them interchangeably.  

Our purpose in this section is to consider the {\it overall} effectiveness of a two-step protocol, but as yet we have only considered the second, measurement, step.  What is the impact of the first, parametric amplification, step? This is modelled by the squeezing Hamiltonian of \erf{hamiltonian}, and, for concreteness, we take $\theta=0$, so that the $X$ quadrature is squeezed. The oscillator also resides in a thermal bath, so the dynamics are governed by a Markovian master equation (ME) consisting of a quadratic Hamiltonian and linear Lindblad (decoherence) operators. Although it is true that this ME maintains the Gaussinity of an initial Gaussian state, we note that the dynamics of the quadrature means (denoted by $X,Y$) and variances ($V_{X}, V_{Y}$) are of a closed form for {\it arbitrary} initial states. The reader should take care to be clear concerning the difference between variances of the system state at time t ($V_{X}, V_{Y}$) and the variance in the estimate of the initial state that exists after some period of measurement ($\sigma_{X}^2,\sigma_{Y}^2$), with the latter being monotonic decreasing in time, as knowledge is acquired. The equations of motion for the $Y$ quadrature of the mechanical oscillator, in the presence of squeezing and thermal bath, but no measurement, are of a linear form for arbitrary initial state,
\bqa
\dot{Y}&=&\frac{1}{2}\left(\chi-\gamma\right)Y\label{meanDot}\\
\dot{V}_{Y}&=&\left(\chi-\gamma\right)V_{Y}+\frac{\gamma}{2}\left(1+2n_{{\rm th}}\right),
\eqa
and can be solved using the initial conditions ($X_{0},Y_{0}$ and $V_{Y}(t=0)=V_{0}$) to give
\bqa
Y&=&Y_{0}e^{(\chi-\gamma)\tau/2}\label{meanY}\\
V_{Y}&=&\left(V_{0}+\frac{\gamma(1+2n_{{\rm th}})}{2 (\chi -\gamma )}\right)e^{ (\chi -\gamma )\tau}
-\frac{\gamma(1+2n_{{\rm th}})}{2 (\chi -\gamma )},
\label{varY}
\eqa
where squeezing for a period of time $\tau$ has been considered.
Note that for squeezing aligned with the $X$ quadrature, the equations of motion for $X,Y$ are uncoupled and that if the covariance between $X$ and $Y$ is initially zero then it will stay zero (as for a coherent state).  It is clear, from these equations, that if $\chi>\gamma$ then there is exponential growth of the $Y$ quadrature mean and variance.  This corresponds to the self oscillation threshold, $\chi_{{\rm osc}} =\gamma $, of a parametric amplifier.  The equations for the $X$ quadrature can be obtained with the replacement $\chi\rightarrow -\chi$, so it follows that if $Y$ is exponentially growing, then $X$ is exponentially damped.

Let us now combine the parametric amplification and measurement steps sequentially, and consider an initial coherent state (which is without loss of generality as has been previously discussed).
Although, the measurement step alone can be described in terms of an effective heterodyne efficiency $\eta_{{\rm het}}$, the overall protocol can not, as the symmetry between quadratures is broken by the squeezing.  The quadrature of most interest is the $Y$ quadrature, as this will contain the bulk of the tomographic information.  After a period, $\tau$, of squeezing, the variance of this quadrature is given by \erf{varY}, but the heterodyne measurement step adds excess noise, of size $\frac{2- \eta_{{\rm het}}}{2\eta_{{\rm het}}}$, to the measured random variable, $y$ (which will be Gaussian if the initial state is Gaussian).
The important point is that the value of $y$ is correlated with that of $Y_{0}$ and we use it to obtain the best estimate of the initial coherent state quadrature, $Y_{{\rm est}}=ye^{-(\chi-\gamma)\tau}$, which has the POVM \erf{eq:POVM} and variance
\beq
\sigma_Y^2=\frac{\chi+2\gamma n_{{\rm th}}}{2(\chi-\gamma)}+
e^{-(\chi-\gamma)\tau}\left(\frac{2- \eta_{{\rm het}}}{2\eta_{{\rm het}}}-
\frac{\gamma(1+2 n_{{\rm th}})}{2(\chi-\gamma)}
 \right).
\label{eq:sigYGen}
\eeq
The expression for $\sigma_X^2$ is once again obtained from that of $\sigma_Y^2$ by the replacement $\chi\rightarrow -\chi$.

A number of results are immediately obtained from \erf{eq:sigYGen}.  These can be summarized by associating with $\chi$ a hierarchy of thresholds
\beq
\chi_{{\rm het}}=2\gamma(1+n_{{\rm th}})\geq\chi_{{\rm del}}=2\gamma\frac{1+n_{{\rm th}}\eta_{{\rm het}} }{2-\eta_{{\rm het}}}\geq\chi_{{\rm osc}}=\gamma. \label{thresholds}
\eeq
The lowest threshold $\chi_{\rm osc}$ is the parametric oscillation threshold.  If we are squeezing for a long time with strength $\chi>\chi_{{\rm osc}}=\gamma$ then a non-zero amount of information will be gathered about the $Y$ quadrature, irrespective of the efficiency of the measurement step.  In contrast, below this threshold, squeezing for a long time leads to essentially no information being gathered since the system approaches a steady state that is independent of the initial condition.  This is illustrated in \frf{analyticComparison2stepSimp} by comparison of the red vertical line, representing $\chi_{{\rm osc}}$, and the dashed black line, representing the $Y$ quadrature estimate variance of the two-step protocol.  In \frf{analyticComparison2stepSimp} the long squeezing duration limit is taken for the two-step protocol.  The middle threshold, $\chi_{\rm del}$, is that required to make squeezing worthwhile as compared to immediately performing the heterodyne measurement with no squeezing step. In other words, $\chi_{{\rm del}}$ is the squeezing strength required to justify delaying the measurement by prolonging the squeezing. To see this, note that if  $\chi>\chi_{{\rm del}}$, then the second term of \erf{eq:sigYGen} is positive, so that it is then beneficial to exponentially suppress it by squeezing for a long time $\tau$.  Alternatively, one can take the partial derivative of \erf{eq:sigYGen} with respect to $\tau$ and note that it is strictly negative for $\chi>\chi_{{\rm del}}$, independent of $\tau$.  The delay threshold, $\chi_{{\rm del}}$, is seen in  \frf{analyticComparison2stepSimp} as the cyan line, which marks the squeezing at which the two-step protocol beats inefficient heterodyne detection.

We are now in a position to claim that for $\chi>\chi_{{\rm del}}$ and the described two-step protocol, the anti-squeezing of the $Y$ quadrature makes it more resolvable.  Conversely, the squeezing of the $X$ quadrature makes it more difficult to probe.  The trade-off between the $X$ and $Y$ quadrature information is not a zero-sum exercise with regards to tomographic performance --- we will show later, when performing tomography on simulated data, that imprecision in quadrature estimates can require an exceedingly large number of samples to be overcome.  Thus, the improved precision achieved in the anti-squeezed quadrature can be very beneficial.  

Just because squeezing is worthwhile does not mean that the heterodyne limit of unity for $\sigma_Y^2$ can be surpassed.  It is only when $\chi>\chi_{{\rm het}}=2\gamma(1+n_{{\rm th}})$ that $\sigma_Y^2\leq 1$.  This can be found from the first term of \erf{eq:sigYGen}. 
For $\chi\gg \chi_{{\rm het}}$, we approach the homodyne limit $\sigma_Y^2\rightarrow \frac{1}{2}$ for a sufficiently long duration of squeezing.   A very important point is that any amount of noise introduced by the measurement step, and captured in the parameter $\eta_{{\rm het}}$, can be suppressed by the exponential.  The $\chi_{{\rm het}}$ threshold is seen in \frf{analyticComparison2stepSimp} as the magenta line and it marks the squeezing at which the two-step protocol enters the homodyne regime.  

Note that none of these thresholds depend on the optomechanical coupling $\mu$; all that is required for a lengthy period of squeezing to lead to essentially ideal homodyne measurement is that the parametric drive $\chi$ can be made sufficiently large compared to the rate at which phonons decohere the oscillator $\gamma (n_{\rm th}+1)$.

In practice there will be other competing effects that would preclude waiting indefinitely during the parametric amplification phase before initiating the measurement. One of these is that the parametric amplification phase, $\theta$, 
will never be perfectly aligned with the quadrature axes defining the measurement --- this will analysed in a later section.  Additionally, there will be mechanical non-linearities not modelled by the parametric amplification Hamiltonian of \erf{hamiltonian}. Both of these effects will mean that in practice there will be some optimal length of time to apply the initial squeezing.  Squeezing beyond this time becomes deleterious.

\begin{figure}
	\centering
\includegraphics[width=0.95\hsize]{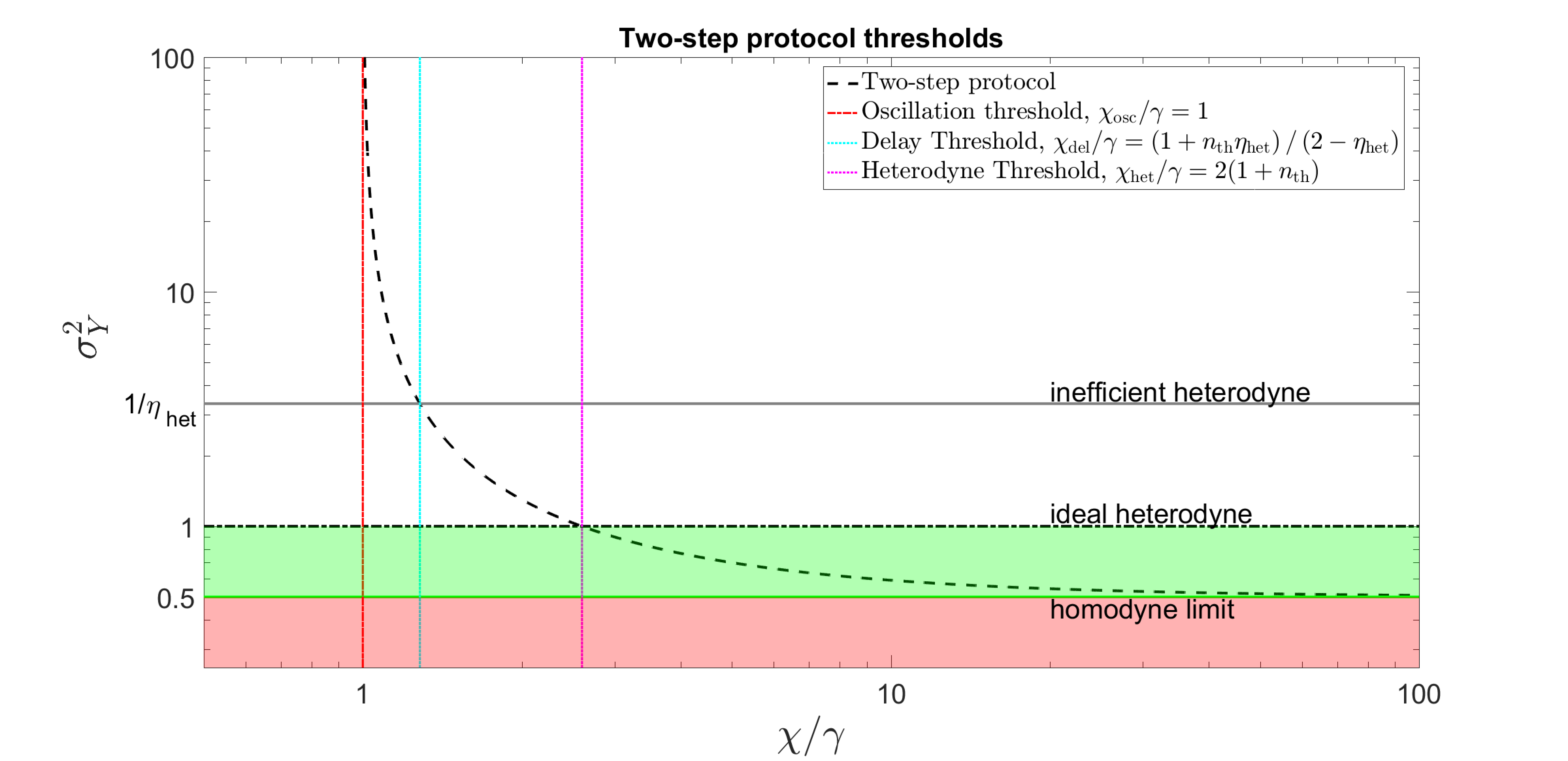}
	\caption{This figure illustrates the three thresholds of parametric amplification with regards to information gathered about the anti-squeezed quadrature.  For convenience, $\chi$ is measured in units of $\gamma$.  The black dashed curve is the variance in the estimate of the initial $Y$ quadrature of a coherent state for the two-step protocol described in the text. The measurement step is modelled as heterodyne detection of some efficiency $\eta_{{\rm het}}$ and the duration of the squeezing step is taken to infinity.  The important thresholds for the parametric amplification strength are shown, the expressions for which can be found in \erf{thresholds}.  Below $\chi_{{\rm osc}}$ (red dash-dot), no information about the initial value of the $Y$ quadrature is obtained.  The parametric drive strength at which a pre-amplification step becomes beneficial is marked by $\chi_{{\rm del}}$ (cyan, dotted). Thus, the horizontal line representing the performance of heterodyne detection with efficiency $\eta_{{\rm het}}$ intersects the curve at this point. For $\chi>\chi_{\rm het}$ (magenta, dotted) the two-step protocol outperforms ideal ($\eta_{\rm het}=1$) heterodyne detection and reaches the homodyne limiting variance of $1/2$ for $\chi\gg\chi_{{\rm het}}$.  To highlight the asymptotic behaviour a log-log scale plot has been used.  The choice of parameters was $n_{{\rm th}}=\eta_{\rm het}=0.3$.
} 
	\label{analyticComparison2stepSimp}
\end{figure}

As it represents one of the major results of this paper, we pause to emphasize the role of $\chi_{{\rm het}}$.  In the absence of a squeezing step, we have explained how the measurement step leads to a heterodyne type measurement of the mechanical oscillator quadratures.  We have then incorporated a parametric amplification step and considered the POVM of the overall protocol, which can be summarized by the value of $\sigma_Y^2$, being the variance of an estimate of the initial $Y$ quadrature of a coherent state.  To highlight the utility of \erf{eq:sigYGen}, it can be seen that it holds true in the case for which the duration of the squeezing step is taken to zero.  In this case, we are returned to $\sigma_{{\rm X}}^{2}=\sigma_{{\rm Y}}^{2}=1/\eta_{\rm het}$.  For non-zero $\tau$, as $\chi$ is increased $\sigma_Y^2$ monotonically decreases (and $\sigma_X^2$ monotonically increases).  This asymmetry in the amount of information gathered about the $X,Y$ quadratures is indicative of the overall measurement protocol now possessing homodyne characteristics.  That is, although the individual measurement step is of a heterodyne type, the composite protocol that includes squeezing has a tomographic performance that has homodyne features. In fact, we can achieve nearly perfect efficiency homodyne limited tomographic performance despite the presence of a finite temperature thermal bath and detector inefficiency.   Physically, when the squeezing is large enough --- that is, when $\chi>\chi_{{\rm het}}$ --- we have found that the amplification dominates the thermal noise and the decohering bath can be effectively frozen out.


To assess the tomographic performance of the two-step protocol, which includes squeezing, it is useful to introduce an effective {\it homodyne} efficiency $\eta_{\rm hom}$.  This is the efficiency of standard homodyne detection that would lead to a variance in the estimate of an initial coherent state quadrature equal to that of our protocol.  The quadrature used in the comparison is that which is probed most closely in our protocol (the $Y$ quadrature in the above discussion) and, of course, the measured homodyne quadrature.  Converting \erf{eq:sigYGen} into an expression for the effective homodyne efficiency is done via $\sigma_Y^2=1/2\eta_{\rm hom}$, which, for $\chi>\chi_{{\rm osc}}$ and $\tau\rightarrow\infty$, gives
\beq
\eta_{{\rm hom}}=\frac{\chi-\gamma  }{\chi +2\gamma  n_{\rm th}}.
\label{homEtaNew}
\eeq
It is clear that perfect efficiency homodyne detection, $\eta_{{\rm hom}}\rightarrow 1$, can be achieved in the large squeezing limit.  Also, note that the details of the measurement step are completely absent --- the protocol has removed all dependence upon the effective heterodyne efficiency and the optomechanical coupling.  This further justifies an initial analysis bereft of the intra-measurement dynamics.

%
%
%
%
%

The simplicity of the form of the parametric amplification strength threshold required to beat the heterodyne limit, $\chi_{{\rm het}}$, invites a heuristic interpretation.  To provide this, we
note that the heterodyne limit can be considered from the point of view of being able to distinguish two coherent states that have a difference in amplitude of unity, despite an additional unit of ground state noise being added due to the measurement of joint quadratures.  Said differently, we can reliably distinguish two equal variance Gaussian distributions when their means differ by more than one standard deviation.  Consider the two Gaussians that we wish to distinguish as being subjected to a long period of squeezing via \erfs{meanY}{varY}.  The ratio of the difference in their means to their standard deviation is given by  $\sqrt{\frac{\chi +2\gamma  n_{\rm th}}{2(\chi-\gamma)  }}$, which is larger than unity for $\chi>\chi_{{\rm het}}=2\gamma(1+n_{{\rm th}})$.  Thus, the squeezing is making coherent states separated by unity more distinguishable and we can beat the heterodyne limit.




To summarize this section, and prepare for the next, we have considered a two-step protocol in which the measurement step was modelled as being heterodyne detection, of efficiency $\eta_{\rm het}$.  A period of parametric amplification constituted the first step and broke the symmetry regarding the variance of estimates of the $X,Y$ quadratures.  This asymmetry made it natural to introduce an overall homodyne effective efficiency to describe the two-step protocol in comparison to standard homodyne detection. The main conclusion was that with sufficiently large squeezing it was possible to approach the limit of ideal homodyne measurement of the initial mechanical state. In the following sections we perform a similar analysis, this time including  the details of the optomechanical measurement in both the zero and blue-detuned laser regimes. Such a full analysis leads to detailed expressions for $\eta_{\rm het}$ in terms of the parameters of the optomechanical system but doesn't alter the qualitative conclusions of this section.   
However, before proceeding with these specific analyses, let us consider the impact of misaligned parametric amplification within the simplified framework of this section. 

\subsection{Misaligned parametric amplification}

Before concluding this section, we investigate the effect of a non-zero value of the parametric drive phase $\theta$, which is inevitable to some extent in practice.  In an experiment, this phase corresponds to the difference between mechanical parametric amplification phase and the phase of the signal used to mix-down the measured photocurrent into a quadrature signal. This relative phase must be controlled or chosen in post-processing, which necessarily involves some uncertainty and fluctuations.  
  Given that the squeezed quadrature estimate variance goes to $\infty$ at long times, it is of concern that any finite misalignment may destroy the desired effect. We can study this issue in the framework of our qualitative analysis where the measurement step is modelled as heterodyne detection with efficiency $\eta_{{\rm het}}$.

The complication, of non-zero $\theta$, couples the equations of motion for the quadrature means and variances, but we can largely avoid this by working in a rotated frame aligned with the squeezing axes.  Then it becomes a matter of transforming the bivariate Gaussian distribution back to the frame defined by the measured quadratures, before adding the diagonal covariance due to measurement inefficiency.  After some algebra, we obtain the variance in the estimate of the initial $Y$ quadrature (we keep in mind that $\theta$ will be small so that the $Y$ quadrature is of interest):
\bqa
\fl\sigma_Y^2=&\frac{\chi+2\gamma n_{{\rm th}}}{2(\chi-\gamma)}\cos^{2}\left(\theta/2\right)+
e^{-(\chi-\gamma)\tau}\left(\frac{2- \eta_{{\rm het}}}{2\eta_{{\rm het}}}-
\frac{\gamma(1+2 n_{{\rm th}})}{2(\chi-\gamma)}
 \right)\cos^{2}\left(\theta/2\right)\nonumber\\
\fl &+\frac{\chi-2\gamma n_{{\rm th}}}{2(\chi+\gamma)}\sin^{2}\left(\theta/2\right)+
e^{(\chi+\gamma)\tau}\left(\frac{2- \eta_{{\rm het}}}{2\eta_{{\rm het}}}-
\frac{\gamma(1+2 n_{{\rm th}})}{2(\chi+\gamma)}
 \right)\sin^{2}\left(\theta/2\right).
\label{eq:sigYGenTheta}
\eqa
From this expression, it is clear that if the squeezing is carried on indefinitely then the variance will exponentially diverge due to the new terms introduced for non-zero $\theta$.  There exists an optimal squeezing duration that can be found by differentiating with respect to $\tau$,
\beq
\fl\tau_{{\rm opt}}=\frac{1}{ \chi }\log \left[\frac{ (2-\eta_{{\rm het}} ) \chi -2 \gamma  (1+\eta_{{\rm het}}  n_{{\rm th}})}{(2-\eta_{{\rm het}} ) \chi + 2\gamma  (1+\eta_{{\rm het}}  n_{{\rm th}})}|\cot\left(\theta /2\right)|\right]
\begin{tikzpicture}
\draw [->] (-0.5,0.1) -- (1.15,0.1);
\draw [white] (0,0) -- (0,0);
\node at (0.3,0.3) {\tiny $\chi\rightarrow\infty, \,\theta\rightarrow 0$};
\end{tikzpicture}
\frac{1}{ \chi }\log \left[\frac{2}{|\theta |}\right].
\eeq
This explicitly shows how sensitive the system is to off-axis squeezing as the optimal time only diverges as $\ln(1/\theta)$ --- unless $\theta$ is extremely small, the optimal time will not be large compared to the other system time scales.  This effect is exacerbated by the $1/\chi$ scaling, with large $\chi$ being the regime of interest.  From \erf{eq:sigYGenTheta}, once this optimal time is exceeded the estimate performance exponentially deteriorates with $\tau$ at a rate $\gamma+\chi$.  This makes it a crucial experimental consideration that may not have been treated appropriately without the above analysis.  Note that a finite squeezing time is optimal even to zeroth order in the bath coupling, showing that it is unavoidable if squeezing off-axis.  If the squeezing is carried out for the optimal duration, $\tau_{{\rm opt}}$, then a term linear in $|\theta |$ is added to the $Y$ quadrature estimate variance.

\section{Full Analysis of Optomechanical Tomography --- Zero-Detuning}
\label{tomogram}

In this section, we provide explicit POVM expressions for the composite optomechanical measurement that arises in the bad-cavity zero detuning regime and also investigate a number of different protocol variations. Unlike the previous section, the details of the measurement step are examined.  We consider both the presence and absence of parametric drive. We are aiming to assess the usefulness of the measurement for tomography, 
 for which accurate estimates of the initial state quadratures must be obtained.  The tools that we allow the experimentalist are continuous measurement and resonant parametric amplification, as illustrated in \frf{schematic}.  In the previous section we found that the effect of the parametric drive was to combat the added noise due to the thermal bath and also to transform the effective measurement for the entire protocol from a heterodyne to homodyne type.  However, it is a priori unclear how the squeezing should be optimally applied and a number of obvious questions can be posed.  Noise both from measurement backaction and from the thermal bath 
 competes with the squeezing, so should the measurement be performed from the start or delayed until the parametric amplification has taken effect?  Do these considerations impact the squeezing threshold $\chi_{\rm het}$ from \erf{thresholds}, that defines when the heterodyne limit can be breached?


To investigate these questions fully, we consider the following scenarios: a two-step process whereby measurement is delayed until the second step --- this is of the same form as the protocol used in our simplistic analysis in the previous section --- and a one-step process in which both measurement and parametric amplification are turned on concurrently.  For the two-step process, we also investigate whether squeezing should be continued in the second of the two steps.  The blue-detuned regime is considered later, in \srf{BlueDetuned} --- in this section we consider procedural variations all with zero-detuning.  A schematic showing how the measurement strength and amplification is varied over the course of the experiment for the one and two-step procedures is shown in \frf{schematic}(b-d).

In contrast to the previous section, we need a stochastic master equation that describes the statistics of the measurement and the state of the mechanical system conditioned on measurements~\cite{WisMil10}. In this zero-detuning, bad-cavity regime, we use the stochastic master equation model of~\cite{PhysRevLett.107.213603,sme}. The two quadratures of the mechanical quantum state are simultaneously probed, see \frf{schematic}(a).  In this regime, the spectral content of the measured signal is concentrated close to $\omega_m$ and the original measurement record, $I(t)=dQ/dt$, is demodulated into quadrature components $I_{X}=dQ_X/dt$ and $I_Y=dQ_Y/dt$~\cite{sme,bowMil} that contain useful information only at low frequencies. The stochastic master equation, as given in \ref{qtraj}, see also~\cite{PhysRevLett.107.213603}, shows that the effect of inefficient detection ($\eta<1$) is to increase the effective temperature of the thermal bath and decrease the effective measurement strength. Therefore certain expressions will be more compact if we define the parameters $n\equiv n_{{\rm th}}+\mu(1-\eta)/\gamma$ and $\tilde{\mu}\equiv \eta\mu$, however, we choose to make the dependence upon $\eta$ explicit in our final expressions, for ease of understanding.

Our approach in the following subsection is to obtain analytic expressions for the variance of the estimates of the initial quadratures,
$\sigma_X$ and $\sigma_Y$, in the case that the input state is a coherent state.
We can use this information to determine an effective detector efficiency for the POVM implemented by the composite measurement. For the moment, it is assumed that the tomographic performance of a measurement scheme 
improves when it becomes more homodyne-like, as represented by increasing effective homodyne efficiency. 
This will be confirmed in \srf{sims}, via simulation of experimental data, where we reconstruct the full quantum state for non-Gaussian inputs.

\subsection{Two-step process}

We begin with the two-step protocol, consisting of a squeezing step that is followed by a measurement step, in which the squeezing is turned off, as this will be familiar to the reader from \srf{simple}.  This is schematically shown in \frf{schematic}(b), with the duration of the measurement step being $T$.  To specify this protocol in detail, all that remains, with regards to the estimate variance of the quadrature of an initial coherent state, is to give the form of $\eta_{{\rm het}}$ that appears in \erf{eq:sigYGenTheta}.  From the POVM (see \ref{qtraj} and~\cite{warDoh}) we find
\beq
\fl\eta_{{\rm het}}=\frac{8 \eta\mu }{\gamma+4 \eta\mu+\Gamma \coth (\Gamma T/2)}
\begin{tikzpicture}
\draw [->] (0.0,0.1) -- (1,0.1);
\draw [white] (-0.2,0) -- (1.3,0);
\node at (0.5,0.3) {\tiny $T\rightarrow\infty$};
\end{tikzpicture}
\frac{8 \eta\mu}{\gamma +4 \eta\mu+\sqrt{\gamma^2+8\eta\mu  (\gamma  +2\gamma  n_{{\rm th}}+2\mu )} },
\label{hetEta}
\eeq
where we have defined the rate $\Gamma = \sqrt{\gamma^2+8\eta\mu  (\gamma  +2\gamma  n_{{\rm th}}+2\mu )}$ and then taken the long measurement time limit, $T\rightarrow\infty$. It is important to note that in the limit of $\gamma, \gamma n_{\rm th}\rightarrow 0$, the effective heterodyne efficiency $\eta_{{\rm het}}\rightarrow 2\eta/(\eta+\sqrt{\eta})\neq \eta$.  This is because $\eta$ refers to efficiency of the actual detector 
in the experimental arrangement of \frf{schematic}(a), rather than the effective efficiency taking into account the strength of the optomechanical coupling and other imperfections (eg the bath). Note that in the regime of $\mu\gg \gamma,\gamma n_{\rm th}$ (or expressed in terms of the optomechanical cooperativity $C\gg 1, \gamma n_{\rm th}$) and $\eta\rightarrow 1$ we have $\eta_{{\rm het}}\rightarrow 1$ so that in the limit of very strong optomechanical coupling and ideal detection the optomechanical measurement step approaches ideal heterodyne detection, as might be expected. We observe that this is also exactly the regime required to feedback cool the oscillator to its ground state~\cite{sme,bowMil}.  From \erf{hetEta} it can be simply shown that the effective efficiency is a monotonic function of each of $\eta, \mu,T,\gamma,n_{{\rm th}}$; we highlight that $\eta_{{\rm het}}$ is monotonic increasing with the measurement strength, $\mu$.  The value of $2/\Gamma$ gives the time scale for the duration of the measurement.  The long time limit, $T\rightarrow\infty$, is very important as it represents a complete measurement with regards to the initial system --- there is no more information available and the final state is no longer correlated with the initial state.

For clarity, and to facilitate comparison with future expressions, the variance of the estimate of the initial $Y$ quadrature is provided in the long measurement time limit, $T\rightarrow\infty$:
\beq
 \fl\sigma_{Y}^2=\frac{\chi+2\gamma  n_{\rm th} }{ 2(\chi- \gamma) }+
e^{- (\chi-\gamma ) \tau } \left(\frac{\gamma
+\sqrt{\gamma ^2+ 8\eta\mu ( \gamma  +2  \gamma  n_{{\rm th}}+2\mu)}}{8\eta\mu}
-\frac{\gamma(1+2 n_{{\rm th}})}{2(\chi-\gamma)}
\right).
\label{2step2}
\eeq
This expression for the variance possesses all the properties described in \srf{simple} (we have merely provided the form of $\eta_{{\rm het}}$) and summarized in \frf{analyticComparison2stepSimp}; in particular, the heterodyne limit of unity can be breached for $\chi>\chi_{{\rm het}}=2\gamma(1+n_{{\rm th}})$ for long squeezing times $\tau$.  As discussed in \srf{simple}, we wish to compare the total measurement protocol to homodyne detection, and in the regime $\chi>\chi_{{\rm osc}}$ and $\tau\rightarrow\infty$ we obtain an effective homodyne detector efficiency
\beq
\eta_{{\rm hom}}=\frac{\chi-\gamma  }{\chi +2\gamma  n_{\rm th}},
\label{homEtaNew2}
\eeq
which approaches unity for large $\chi\gg\chi_{{\rm het}}$.  The reader will recall that this homodyne efficiency is found by comparing the variance of the anti-squeezed quadrature to that of the measured quadrature in standard homodyne detection.  It is the anti-squeezed quadrature that has the improved resolution, while we lose the ability to learn about the squeezed quadrature.

Although our primary concern in this section is to analyse tomographic performance, it is important to identify how one actually arrives at the initial quadrature estimate.  That is, we now provide the filter on the measurement results that best estimates the state we wish to reconstruct.  For display brevity, we take the long measurement time limit and find
\beq
Y_{\rm est}(T)
\begin{tikzpicture}
\draw [->] (0.0,0.1) -- (1,0.1);
\draw [white] (-0.2,0) -- (1.3,0);
\node at (0.5,0.3) {\tiny $T\rightarrow\infty$};
\end{tikzpicture}
\frac{1}{4\sqrt{\mu \eta }}\left(\gamma +\Gamma\right)
e^{- (\chi-\gamma )\tau/2} \int_{0}^{T}e^{-\Gamma t/2 }dQ_Y (t),
\label{filter}
\eeq
with the expression for $X_{\rm est}$ obtained with $\chi\rightarrow-\chi$.  The reader will note that there are two aspects to the filter: the system state at the conclusion of the squeezing is estimated based on the measurement results, $dQ_{Y}$, and, secondly, the deterministic pre-image of this state is found by inverting the squeezing (and thermal bath) evolution of \erf{meanY} (represented by the $e^{- (\chi-\gamma )\tau/2} $ term).  This can be compared with \erf{filterGeneral} (using $dQ_{Y}=dtI_{Y}$) to discern the filter function, $h_Y(t)$.  We see that the early measurement times (up to a time $2/\Gamma$) contribute the most to our estimate and that the contributions die off exponentially. 

We have seen that a squeezing step can be beneficial to the precision of our estimate of the quadrature that is amplified, but only when $\chi>\chi_{{\rm del}}$.  Below this limit, the noise from the thermal bath overcomes the amplification effect during the pre-measurement phase (squeezing).  However, for a fixed squeezing duration, increasing $\chi$ improves performance even below $\chi_{{\rm del}}$.  It is logical to then suggest that the squeezing should be maintained during the measurement step also; that is, throughout the entire protocol, as per \frf{schematic}(c).  This has the effect of transforming the measurement step away from a purely heterodyne form.  It is then no longer sensible to ascribe an effective heterodyne efficiency to the measurement step, but there is no difficulty in determining the POVM, using the techniques of \cite{warDoh}.  Assuming that the $X$ quadrature is squeezed with a constant amplitude, with the measurement turned on at time $\tau$ for a duration $T$ leads to the POVM variance
\beq
\fl\sigma_{Y}^2=\frac{\chi+2\gamma  n_{\rm th} }{ 2(\chi- \gamma) }+e^{-(\chi-\gamma )   \tau } \left(
\frac{\Gamma_{-} \coth (\Gamma_{-}T/2)+\gamma- \chi }{8 \eta\mu}
-\frac{\gamma(1+2 n_{{\rm th}})}{2(\chi-\gamma)}
\right),
\label{2stepmeas2}
\eeq
with the rates defined as $\Gamma_{\pm} = \sqrt{(\gamma\pm \chi)^2+8\mu\eta  (\gamma  +2\gamma   n_{{\rm th}}+2\mu )}$.  The expression for $\sigma_{X}^2$ requires $\chi\rightarrow-\chi$ and, resultantly, involves the rate $\Gamma_{+}$.  If we set $\chi=0$ then $\Gamma_{\pm}=\Gamma$, which is as expected as this returns us to the previously described protocol of \frf{schematic}(b).
It is clear that we can, once again, breach the heterodyne limited variance of unity, and enter the homodyne paradigm, for $\chi>\chi_{{\rm het}}$, as this allows the exponential to be suppressed, for long squeezing, and the first term to drop below unity.  It is straightforward to show, by comparison of \erf{2step2} and \erf{2stepmeas2} in the long measurement time limit, $T\rightarrow\infty$, that maintaining squeezing in the second step leads to a smaller (or equal) estimate variance in all cases.  It is therefore evident that the squeezing concurrent with measurement is having a beneficial effect, in addition to the pre-measurement squeezing.  Rather than proceeding with the analysis in the context of a two-step protocol, it is more enlightening to isolate its effect further and reduce to a one-step protocol.  That is, we remove the pre-measurement squeezing step and consider the protocol of \frf{schematic}(d), where both measurement and squeezing are (always) performed together. 

\subsection{One-step process}
The one-step protocol (see \frf{schematic}(d)) POVM variance is most simply obtained by reducing the duration of the squeezing step of the two-step protocol to zero ($\tau\rightarrow 0$ in \erf{2stepmeas2}):
\bqa
\sigma_{Y}^2&\quad=&\frac{1}{2}+\frac{\gamma -\chi +\Gamma_{-} \coth (\Gamma_{-} T/2)}{8 \mu \eta}\nonumber\\
&\begin{tikzpicture}
\draw [->] (0.0,0.1) -- (1,0.1);
\draw [white] (-0.2,0) -- (1.3,0);
\node at (0.5,0.3) {\tiny $T\rightarrow\infty$};
\end{tikzpicture}&
\frac{1}{2}+\frac{\gamma -\chi +\sqrt{(\gamma -\chi)^2+8\mu\eta   (\gamma +2\gamma   n_{{\rm th}}+2\mu )}  }{8\mu\eta }.
\label{sigma}
\eqa
Furthermore, the filter for the single step protocol is
\beq
Y_{\rm est}(T)
\begin{tikzpicture}
\draw [->] (0.0,0.1) -- (1,0.1);
\draw [white] (-0.2,0) -- (1.3,0);
\node at (0.5,0.3) {\tiny $T\rightarrow\infty$};
\end{tikzpicture}
\frac{1}{4\sqrt{\mu \eta }}\left(\gamma - \chi +\Gamma_{-}\right)
\int_{0}^{T}e^{-\Gamma_{-} t/2 }dQ_Y (t),
\label{filter2}
\eeq
which should be compared with \erf{filter}.  The new protocol modifies the time scale to $2/\Gamma_{-}$ and makes the multiplicative constant $\chi$ dependent, together with removing the need to obtain the state's pre-image under pre-amplification (ie there is no squeezing only step).  

It should come as no surprise to the reader that $\sigma_{Y}^2$ is monotonic decreasing with $\chi$, as can be easily verified by calculating $d\sigma_{Y}^2/d\chi$.  In fact, for $\chi\gg\gamma,\gamma n_{{\rm th}},\mu$ the term of ${\cal O}(\chi)$ cancels and we obtain $\sigma_{Y}^2\rightarrow 1/2$.  It is therefore possible to reach the homodyne limit with a single step protocol, as long as the squeezing is powerful enough.  Another obvious limit to take is $\mu\gg\gamma,\gamma n_{{\rm th}},\chi$, which leads to an instantaneous measurement of a heterodyne type (we are taking the limit in which the squeezing is unimportant).  Thus, \erf{hetEta} is applicable and if $\eta\rightarrow 1$ then $\sigma_{Y}^2\rightarrow 1$, with ideal heterodyne detection resulting ($\eta_{{\rm het}}\rightarrow 1$).

We have seen that sufficiently large squeezing leads to the homodyne paradigm, whilst very fast measurement returns us to an instantaneous heterodyne measurement.  In the two-step protocol, very large squeezing was {\it not} required to breach the heterodyne limit, rather, $\chi>\chi_{{\rm het}}=2\gamma(1+n_{{\rm th}})$ was sufficient.  The question remains as to whether there is an analogous regime for the one-step protocol.  Clearly, $\chi>\chi_{{\rm het}}$ with finite $\gamma,n_{{\rm th}},\mu$ will not suffice.  We can guess what is required, however, by using the intuition gained from the two-step protocol.  In that case, the quadrature needed to be sufficiently amplified before it was measured.  If the measurement is applied too soon, then the backaction heating is also amplified, which obscures the information contained in future measurements.  With the measurement and squeezing in a single step, how can we delay the measurement?  The answer is to use a very weak measurement strength, which postpones the effective wavefunction collapse until the amplification has occured.  That is, very weak measurement is a proxy for a delayed measurement, and a pre-measurement amplification is then effectively achieved.

That very weak measurement can be beneficial, and allow the homodyne paradigm to be reached, is confirmed by minimising $\sigma_{Y}^2$ with respect to $\mu$.  
We find that the minimum of $\sigma_{Y}^2$ is obtained either by $\mu\rightarrow 0$ or by $\mu\rightarrow\infty$.  Depending upon whether we are below or  above a squeezing strength threshold, it is better to measure as quickly as possible or to delay as long as possible (indicated here by a very weak measurement).  This threshold is found by noting that the derivative of $\sigma_{Y}^2$ with respect to $\mu$ flips sign when $\chi=\chi_{{\rm del}}=2\gamma (1+n_{{\rm th}})+\gamma(\sqrt{\eta}-1)$. That is, $\sigma_{Y}^2$ is monotonic increasing with $\mu$ when $\chi>\chi_{{\rm del}}$ and monotonic decreasing when $\chi<\chi_{{\rm del}}$.  If we are above this threshold  for $\chi$, then it is optimal to choose $\mu\rightarrow 0$.  
The reader will notice that we have used the same nomenclature, $\chi_{{\rm del}}$, as for the two-step protocol when indicating the threshold at which it is best to delay measurement, whether it be by a longer pre-amplification time, $\tau$ (two-step), or by a weaker measurement strength, $\mu$ (one-step).  Apart from the conceptual similarity, we note their equality when $\eta_{{\rm het}}=2\eta/(\eta+\sqrt{\eta})$ is used in the expression for $\chi_{{\rm del}}$ from the two-step protocol ($2\gamma\frac{1+n_{{\rm th}}\eta_{{\rm het}} }{2-\eta_{{\rm het}}}$).  This value of $\eta_{{\rm het}}$ corresponds to an instantaneous measurement, in the large $\mu$ limit, for our optomechanical system.  The reason for the coincidence of $\chi_{{\rm del}}$ for both protocols arises due to the fact that the same limiting performance is achieved when the measurement strength is optimally chosen.

If the measurement is delayed using an infinitesimal measurement strength then, when we are above $\chi_{{\rm osc}}=\gamma$, the $Y$ quadrature estimate variance is given by 
\beq
\sigma_{Y}^2
\begin{tikzpicture}
\draw [->] (-0.7,0.1) -- (1.34,0.1);
\draw [white] (0,0) -- (1.6,0);
\node at (0.3,0.3) {\tiny $\chi>\gamma \enskip{\rm and}\enskip \mu\rightarrow 0$};
\end{tikzpicture}
\frac{\chi +2\gamma  n_{\rm th}}{2( \chi-\gamma ) },
\label{yOpt}
\eeq
which is exactly the same as for the two-step protocol.  Thus, once again, the regime defined by $\chi>\chi_{{\rm het}}=2\gamma(1+n_{{\rm th}})$ corresponds to the minimum $\chi$ required to breach the heterodyne limited variance of unity.  In the limit of large $\chi$ the homodyne bound of $\half$ is reached.  It is important to note that this is true even when the detector is not perfectly efficient.  Finally, if the delayed measurement protocol was chosen for $\chi<\chi_{{\rm osc}}$, then the estimate variance becomes infinite. Consequently we have established a hierarchy of squeezing thresholds for the one-step protocol, just as for the two-step protocol: $\chi_{{\rm het}}\geq\chi_{{\rm del}}\geq\chi_{{\rm osc}}$.  These thresholds have the analogous interpretation as for the two-step protocol, but now with the replacement of a long duration pre-amplification step by an infinitesimal measurement strength.  That is, $\chi_{{\rm osc}}$ represents the threshold for long duration squeezing leading to a well-defined estimate of the initial state, $\chi_{{\rm del}}$ represents the threshold for it being optimal to delay the measurement and $\chi_{{\rm het}}$ being the squeezing value sufficient to enter the homodyne paradigm.  The thresholds for the one-step protocol could be illustrated in a very similar way to \frf{analyticComparison2stepSimp}, as the same threshold structure exists, provided that the infinitesimal measurement strength limit was taken instead of a very long pre-amplification step.



We have found that there are two pathways to the homodyne regime with finite squeezing.  Either a pre-amplification step can be used (giving a two-step protocol) {\it or} squeezing can be turned on concurrent with measurement.  In the former method, long squeezing times are required, while in the latter, very weak measurement is necessary.  Both methods need $\chi>\chi_{{\rm het}}=2\gamma(1+n_{{\rm th}})$.  
Although both one and two-step protocols reach the homodyne limit given freedom to choose $\mu$, which protocol is optimal if the measurement strength is finite and fixed due to experimental reasons? The $\chi_{{\rm del}}$ threshold for the two-step process provides us with the answer to this, essentially by definition, with the two-step (one-step) protocol being preferred when $\chi>\chi_{{\rm del}}$ ($\chi<\chi_{{\rm del}}$).

Further comparison between the one and two-step protocols is made in \frf{analyticComparison}, where we compare the analytic expressions for the one and two-step variance (with squeezing maintained throughout) for zero-detuning for some realistic parameters.  The blue-detuned variance (to be discussed in \srf{BlueDetuned}) is also plotted.  In \frf{analyticComparison}(a) it can be seen that all processes perform better as the squeezing is increased and the threshold squeezing, $\chi_{{\rm het}}$ (magenta line), at which the heterodyne limiting variance of unity is beaten, is approximately confirmed.  The small offset from the threshold  is due to a finite duration of squeezing.
In \frf{analyticComparison}(b), parameters are chosen such that $\chi>\chi_{{\rm het}}$.  The main result is that the one-step processes (dashed lines) degrade with increasing measurement strength, as the squeezing has less time to take hold before the state becomes decorrelated with the initial state.  This effectively reduces the amount of squeezing and leads to reduced resolution, consistent with \frf{analyticComparison}(a).  In the large $\mu$ limit the one-step variance will approach the heterodyne value of approximately one, with a small overshoot due to the finite detector inefficiency.  The two-step processes are less affected by the reduced squeezing in the second step as the first step also provides squeezing, so they will asymptote to a sub unity value.  All curves in \frf{analyticComparison}(b) meet as $\mu\rightarrow 0$.  This is because the variance saturates the bound of $\frac{  \chi+2\gamma  n_{{\rm th}} }{2( \chi -\gamma )}$ in this limit.


\begin{figure}
	\centering
 \includegraphics[width=0.95\hsize]{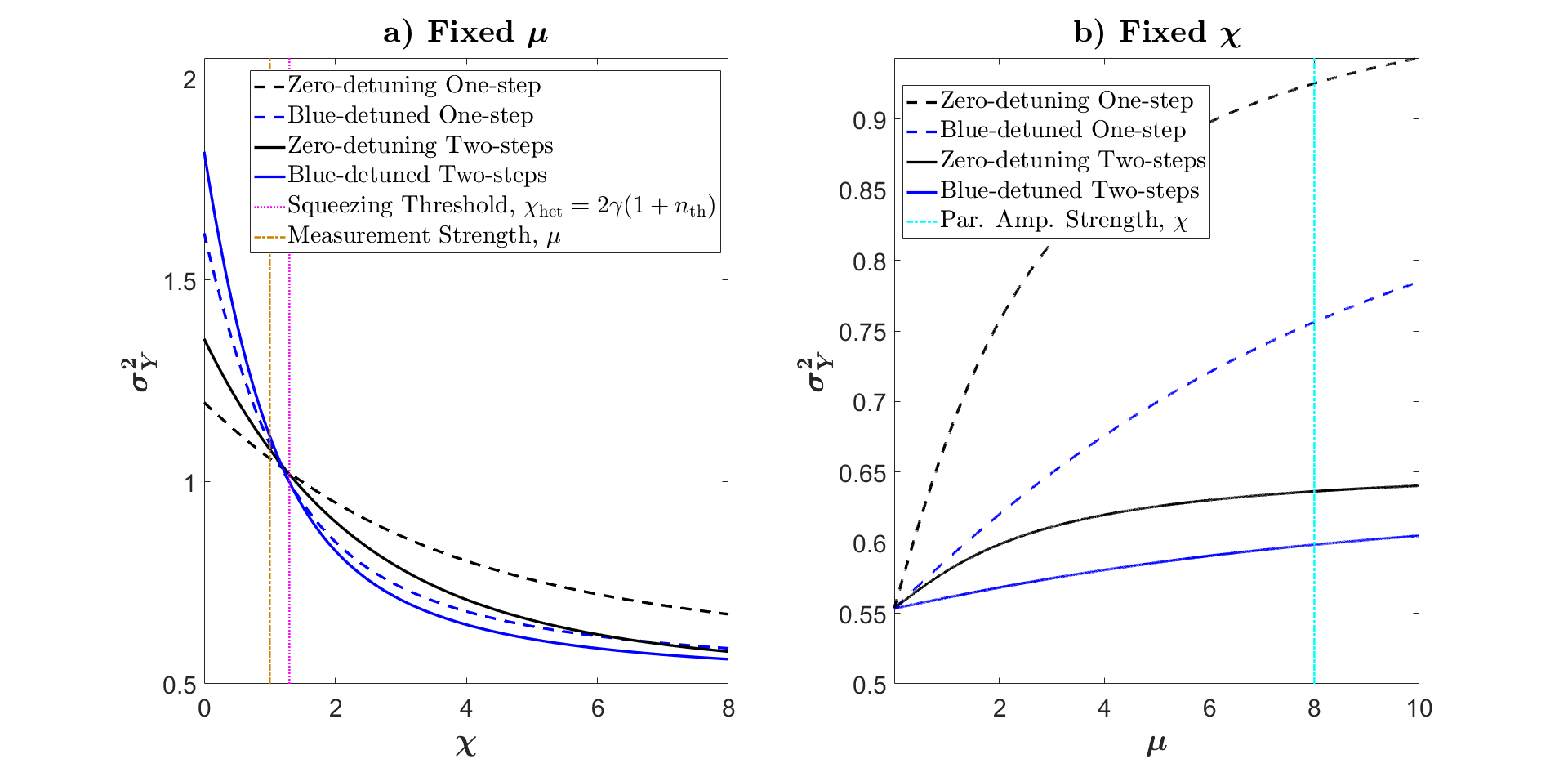}
	\caption{The one-step (dashed) and two-step (solid) $Y$ quadrature estimate variances of an initial coherent state for the zero-detuned (black) and blue-detuned (blue) cases are compared. 
For the two-step protocols, squeezing is maintained in the second step.
{\bf a)} The measurement strength $\mu=1$ is held fixed (shown in brown) while squeezing $\chi$ is varied. Also shown is the threshold to beat the heterodyne limit, $\chi_{{\rm het}}$ (magenta).  For both zero and blue-detuning, the two-step processes outperform the one-step processes when $\chi>\chi_{{\rm het}}$ (to the right of the magenta line).  The variances asymptote towards $\frac{1}{2}$ for very large $\chi$.  Other parameters are: $\gamma=0.5,  \eta=0.9, n_{{\rm th}}=0.3, \tau=0.2, T=5$.  
{\bf b)} Blue and black lines are defined as per (a) but now with $\chi=8$ held fixed (shown in cyan) and varying measurement strength, $\mu$.  As $\chi>\chi_{{\rm het}}$ the variance degrades as $\mu$ increases --- there is less time for the squeezing to amplify the quadrature of interest.  }
	\label{analyticComparison}
\end{figure}

\section{Analysis of Optomechanical Tomography --- Blue-Detuned Laser}
\label{BlueDetuned}
In this section, we repeat the analysis of section~\ref{tomogram}, but now in the blue-detuned resolved sideband limit. As noted previously, this regime can also be described, approximately, using a single mode stochastic master equation.
The derivation of the effective single-mode system involves a rotating wave approximation in the frame of the driving laser, which is blue-detuned by approximately the mechanical oscillator frequency, $\Delta\sim\omega_{m}$~\cite{revCavOptMech}. We work in a regime where $g\ll\kappa\ll\omega_{{\rm m}}$, which has been experimentally implemented~\cite{BDexpI,BDexpII,bdexperiment}.
This regime is of particular relevance because of the amplification provided by the resultant two-mode squeezing interaction between cavity and mechanical oscillator.  The impact of this new source of amplification should be compared with the mechanical squeezing previously considered.  
Further details of the the quantum trajectory methods involved in obtaining the POVM for the blue-detuned system are provided in \ref{qtraj}. 


\subsection{Two-step process}
\label{BlueDetunedTomography}
To investigate the POVM implemented by the composite optomechanical measurement, we again consider the variance of the quadrature estimate of an initial coherent state amplitude and the corresponding effective homodyne efficiency.  That this figure of merit is strongly tied to the ability to reconstruct states will be shown in \srf{sims}.  Consistent with the previous sections, we choose to squeeze the $X$ quadrature in order that the $Y$ quadrature becomes more resolvable.  The first protocol we discuss is the two-step process of \frf{schematic}(b) (no squeezing in the second step), as then the only missing piece from \erf{eq:sigYGen} is the $\eta_{{\rm het}}$ appropriate for the blue-detuned resolved sideband measurement step.

Before providing a general expression that is also a function of the thermal bath, it is illuminating to consider just the effect of detector inefficiency and finite measurement strength, as this exposes differences due to the blue-detuned regime.  That is, we make the simplification $\gamma,n_{{\rm th}}=0$, and isolate the amplifying effect of the detuned laser adding energy to both the optical and mechanical modes.  Considering an incomplete measurement that lasts for time $T$ we find that the effective heterodyne efficiency for the measurement step is
\begin{equation}
\eta_{\rm het} = \frac{2\eta}{2\eta + \coth \left(\mu  T/2\right)-1}.
\end{equation}
Notice that in the limit $T\rightarrow \infty$ the efficiency approaches 1, independent of $\eta$. This is a result of the intrinsic amplification that is available in the blue-detuned case, and should be compared to the zero-detuned case where we obtained $\eta_{\rm het} = 2\eta/(\eta+\sqrt{\eta})$. However, the optical efficiency $\eta$ does still result in a slower rate of information gain, meaning that the time required to achieve a given efficiency $\eta_{\rm het}$ grows as $\eta$ is decreased.  Although the detector losses are overcome by the blue-detuning, this does not transform the protocol into being of an effective homodyne type --- parametric amplification is still required for that.

We now show a heterodyne efficiency for a completed measurement, $T\rightarrow\infty$, with no squeezing applied, but  with non-zero $\gamma,n_{{\rm th}}$:
\begin{equation}
\eta_{\rm het} = \frac{2\eta\mu}{\gamma+\mu(2\eta-1)+\sqrt{(\mu-\gamma)^2+4\eta\mu\gamma(n_{\rm th}+1)}},
\end{equation}    
which can be directly compared to the corresponding expression for the zero-detuned case~(\ref{hetEta}).  This can be substituted into \erf{eq:sigYGen} to obtain an estimate variance for the two-step protocol.  It is clear that the effective {\it homodyne} efficiency for the overall two-step protocol, above parametric oscillation threshold and for long squeezing times, is given by 
\beq
\eta_{\rm hom}
\begin{tikzpicture}
\draw [->] (-0.5,0.1) -- (1.34,0.1);
\draw [white] (0,0) -- (1.6,0);
\node at (0.35,0.3) {\tiny $\chi>\gamma, \, \tau\rightarrow\infty$};
\end{tikzpicture}
\frac{\chi-\gamma }{\chi +2\gamma  n_{{\rm th}}  },
\label{BluexOpt}
\eeq
exactly the same as the zero-detuned case~(\ref{homEtaNew2}).  Ideal homodyne detection is achieved by the two-step protocol in the blue-detuned regime when, as before, $\chi\gg \chi_{{\rm het}}=2\gamma(1+n_{{\rm th}})$.

We now consider maintaining the squeezing amplitude throughout the measurement step (as illustrated in \frf{schematic}(c)).  This leads to a $Y$ quadrature estimate variance of 
\beq
\fl\sigma_{Y}^2=\frac{\chi+2\gamma  n_{{\rm th}} }{2( \chi -\gamma) }+
e^{- ( \chi-\gamma) \blk \tau } \left(\frac{{\it \Gamma}_{-}  \coth \left({\it \Gamma}_{-} T/2\right)+\gamma-\chi+\mu  (\eta-1 )}{2 \eta  \mu}
-\frac{\gamma(1+2 n_{{\rm th}})}{2(\chi-\gamma)}
\right)
\label{BDsigma2StepNew}
\eeq
where we have defined the rates ${\it \Gamma}_{\pm} = \sqrt{(\gamma \pm  \chi )^2\mp 2 \mu  \chi( 1-\eta)  +\mu ^2+2\gamma  \mu  (2 \eta  (n_{{\rm th}}+1)-1)}$, which are of the same order as the zero-detuned rates $\Gamma_{\pm}$.  The filter for this protocol is given by 
\beq
Y_{\rm est}(T)
\begin{tikzpicture}
\draw [->] (0.0,0.1) -- (1,0.1);
\draw [white] (-0.2,0) -- (1.3,0);
\node at (0.5,0.3) {\tiny $T\rightarrow\infty$};
\end{tikzpicture}
c e^{- (\chi-\gamma )\tau/2} 
\int_{0}^{T}e^{-{\it \Gamma}_{-}t /2}dQ_{Y} (t),
\label{BDfilterNew}
\eeq
where the rather lengthy multiplicative constant, $c$, is defined in the footnote \cite{constantBD}.

In \frf{analyticComparison}(a), the estimate variance of the $Y$ quadrature for the two-step blue-detuned protocol, with squeezing maintained throughout, is plotted against squeezing strength, $\chi$ (solid blue).  Compared with the zero-detuned two-step process, blue-detuning performs better when $\chi>\chi_{{\rm het}}$ but worse otherwise.  This is due to the additional amplification that blue-detuning provides.  When  $\chi<\chi_{{\rm het}}$ the noise is dominating the squeezing, so that amplification decreases the signal (the initial state) to noise (the amplified contribution).  When $\chi>\chi_{{\rm het}}$, amplification is desired as it makes the signal more resolvable, so that blue-detuning is superior.  This effect is strong enough that the one-step blue-detuned variance (to be given shortly) is less than the two-step zero-detuned variance for much of the parameter range displayed.  In \frf{analyticComparison}(b), the variation against measurement strength, $\mu$ is shown.  As $\chi>\chi_{{\rm het}}$, blue-detuning outperforms zero-detuning.  The composite measurement degrades when the measurement strength increases as this reduces the effective squeezing provided in the second step.
Although slower than the zero-detuned case, the variance will approach approximately unity for very large $\mu$. 


\subsection{One-step process}
\label{BlueDetunedTomography2}

For completeness, we provide the variance of the $Y$ quadrature estimate for the one-step protocol (see \frf{schematic}(d)), by taking the $\tau\rightarrow 0$ in \erf{BDsigma2StepNew}, to obtain
\bqa
\sigma_{Y}^2&\quad=&\frac{1}{2}+\frac{\gamma-\chi +\mu(\eta-1)+{\it \Gamma}_{-} \coth ({\it \Gamma}_{-} T/2)}{2 \eta\mu }\nonumber\\
&\begin{tikzpicture}
\draw [->] (0.0,0.1) -- (1,0.1);
\draw [white] (-0.2,0) -- (1.3,0);
\node at (0.5,0.3) {\tiny $T\rightarrow\infty$};
\end{tikzpicture}&
\frac{1}{2}+\frac{\gamma - \chi +\mu(\eta -1)  + {\it \Gamma}_{-} }{2 \eta\mu },
\label{BDsigmaNew}
\eqa
with ${\it \Gamma}_{-}$ as per the blue-detuned two-step protocol.  The filter is also obtained from \erf{BDfilterNew} with $\tau\rightarrow 0$.  There is much similarity with previous protocols, in particular the thresholds $\chi_{{\rm osc}}=\gamma$ and $\chi_{{\rm het}}=2\gamma (1+n_{{\rm th}})$ are the same and are once again achieved via weak measurement, $\mu\rightarrow 0$, delaying the decorrelation of the initial system state until the squeezing has had time to take effect.  As for the other protocols we have considered, it it possible to approach ideal homodyne detection with $\eta_{\rm hom}\rightarrow 1$. A difference worth mentioning is that the threshold at which it is best to delay the measurement is the same as that required to enter the homodyne regime; that is, $\chi_{{\rm del}}=\chi_{{\rm het}}$.  The one-step blue-detuned estimate variance is plotted (dashed blue), for comparison, in \frf{analyticComparison}.

\section{Tomographic Reconstruction From Simulated Data}
\label{sims}
Up until this point, we have calculated the POVM implemented by the composite optomechanical measurement, and assumed that the tomographic performance of a measurement scheme improves when the effective homodyne efficiency of the measurement is increased. In converting the optomechanical measurement to an effective homodyne measurement, we have sacrificed accuracy in the squeezed quadrature in order to beat the heterodyne limit in the anti-squeezed quadrature.  Motivating this construction is the idea that tomography will become much more difficult as the accuracy of the quadrature estimate decreases, while reduction in the effective number of data samples (information is gained about one quadrature only) can be overcome with a doubling only of the number of trials.  Similar concepts are discussed in~\cite{tomRevAriano} where detector inefficiency introduces an anti-Gaussianity into estimators for tomography that cause a much slower statistical convergence.  In this section, we justify this assumption by simulating the system of interest. This produces artificial experimental data that can be used as the input to a tomography reconstruction algorithm. 

Not all states are equally difficult to reconstruct --- our scheme will be of greatest benefit when high resolution is important.  For example, the Wigner function of a coherent state can be well approximated from the results of a coarse-grained phase space probe, but the Wigner function of a superposition of coherent states (a `cat' state) requires a finer probe to observe the faster oscillations.  This amounts to an expectation that homodyne rather than heterodyne detection is preferred as it provides more accurate quadrature estimates.  In this section we will investigate this claim by performing Maximum Likelihood (MaxLik) tomography~\cite{tomRevLvovsky} on simulated experimental data.  As we have analytic expressions for the POVM that represent the composite measurement on the system, MaxLik tomography can readily be implemented. This is a standard procedure that is widely implemented in the experimental literature and also has the advantage of producing physical density matrices in contrast to Wigner function tomography~\cite{tomRevLvovsky}. A study of alternative approaches to state reconstruction is beyond the scope of this paper. From our previous analysis, and also \frf{analyticComparison}, it is clear that the (experimentally accessible) two-step procedure outperforms the one-step when $\chi>\chi_{{\rm het}}$, so for this reason it is the focus of the simulations.  The zero-detuned case is considered as the results are expected to carry over to the blue-detuned case, which is of greater difficulty to simulate due to the laser amplification necessitating much larger basis size. 
To highlight the difficulties faced by heterodyne detection, and how these can be overcome with resonant parametric amplification, the input state that we choose to reconstruct is a cat state of the form $\ket{{\rm cat}}\propto\left(\ket{\alpha}+\ket{-\alpha}\right)$.

The experimental procedure that we simulate is the collection of the measurement data $dQ_{X,Y}(t)$ for a large number of repeated trials.  Each trial in the laboratory would consist of state preparation followed by a period of electromechanical driving with a particular phase (step one) and finally the measurement is turned on (step two) and the $dQ_{X,Y}(t)$ are recorded.  The distribution of squeezing phases should be sufficient to measure a representative sample from the continuum of quadratures. Here we choose to simulate an equal number of trials at each of $50$ uniformly distributed phases.  The measurement results are then filtered (data analysis) to determine the POVM element relevant for that trial (see \ref{qtraj}).  The collection of POVM elements, one for each trial, is then fed into the MaxLik algorithm.  The output of the algorithm is the initial state most likely to have produced the set of recorded measurement results.  This state is taken as the best estimate of the initial state.       

The first result that we provide, see \frf{wignerReconstruction}, is a comparison of the Wigner functions of density matrices reconstructed with realistic choices of parameters (see figure caption).  First we show the target state (a), then a reconstruction with no squeezing (b) and progressively more squeezing (c),(d). It is striking that in the absence of squeezing, no Wigner function negativity is detected for the cat state, $\propto\left(\ket{\alpha=3}+\ket{\alpha=-3}\right)$.  This is not a general feature.  A $2$ phonon Fock state for example --- which has Wigner function negativity --- can be well reconstructed with no squeezing due to its slower oscillations.  Our chosen benchmark for comparing the performance of our scheme is optomechanical measurement with no squeezing. In this case it is not sensible to have a step one, which would just be a waiting period during which the initial mechanical state would decohere. Consequently, such a wait period is omitted and the optomechanical measurement amounts to heterodyne measurement of the initial state.  As squeezing is progressively added, the oscillatory features become better resolved.  With the largest simulated value of squeezing, in subplot (d), the features are quite accurately placed, which is not the case in subplot (c).    

\begin{figure}
	\centering
	\includegraphics[width=0.95\hsize]{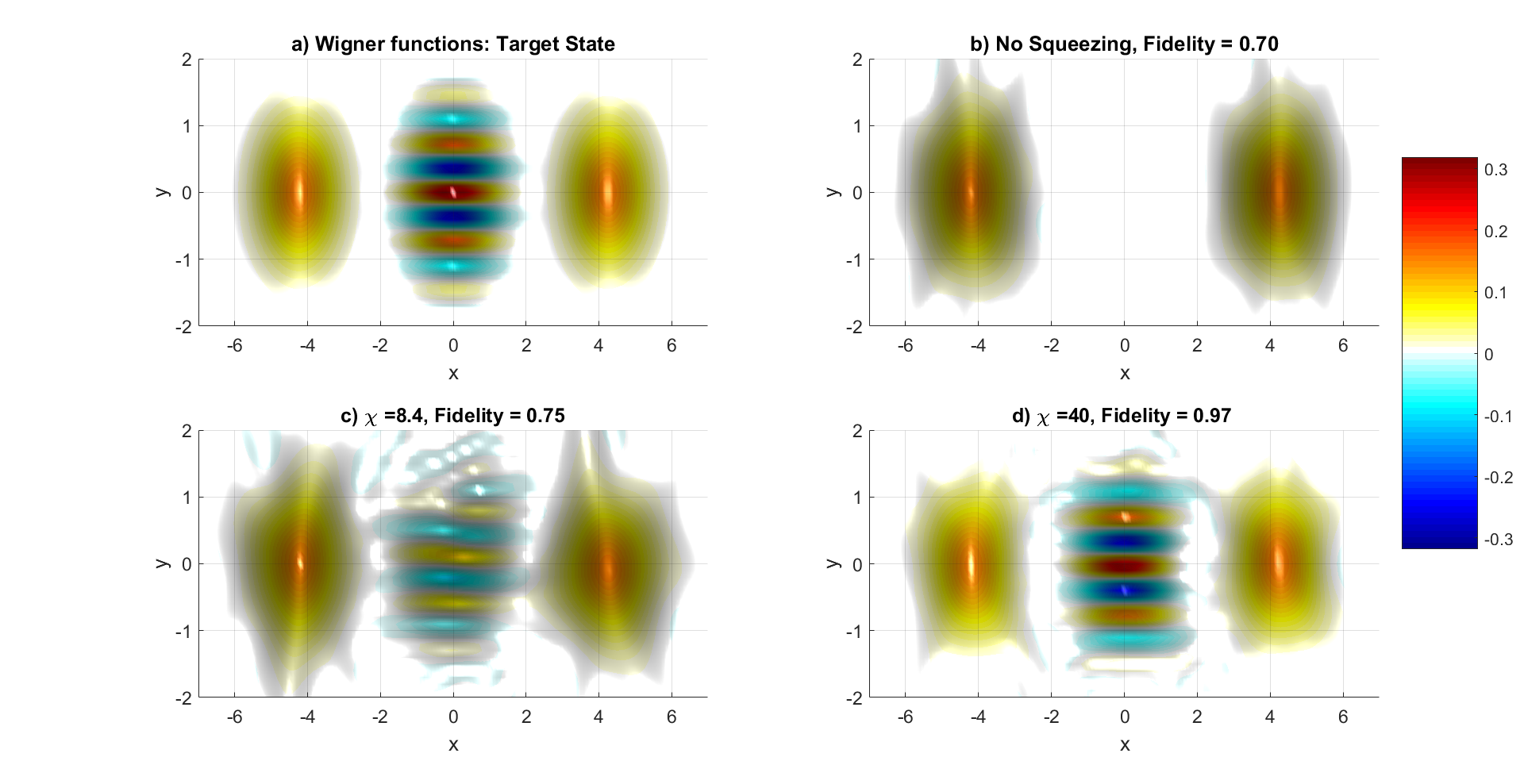}
	\caption{{\bf a)} Wigner function of the cat state ($\propto\left(\ket{\alpha=3}+\ket{\alpha=-3}\right)$).  {\bf b)} Tomographic reconstruction of the state in (a) using inefficient heterodyne detection corresponding to optomechanical position measurement with no squeezing.   In {\bf c)} and {\bf d)} squeezing is added using the two-step protocol and zero-detuning position measurement. In (c) $\chi=8.4$ and in (d) $\chi=40$.  Although subplot (d) has some very low amplitude noise it resolves all features clearly and has a fidelity of $0.97$. Parameters used for the simulations are: $\gamma=0.5, \eta=0.9,n_{{\rm th}}=0.3,\mu=50, \tau=2, T=2$, where $\tau$ and $T$ are quoted in units of the characteristic time scale, $2/\Gamma_{-}$.  4000 trials are used for each reconstruction, with the trials spread evenly across $50$ uniformly spaced squeezing angles (where relevant).
	}
	\label{wignerReconstruction}
\end{figure}

Motivated by the qualitative nature of the Wigner function comparisons, a more detailed investigation of the fidelity metric is completed for an amplitude $\alpha=2$ cat state (note the differing amplitude to the \frf{wignerReconstruction} simulation).  In \frf{fidPlot}, the fidelity of the reconstructed state with respect to the initial cat state is determined as a function of the number of trials, for a range of different squeezing amplitudes.  We also perform tomography with several benchmarks for comparison.  The benchmarks are perfect efficiency homodyne and heterodyne detection, and also, inefficient homodyne ($\eta_{\rm hom}=0.9$) and heterodyne detection ($\eta_{\rm het}=0.97$, which is equivalent to a detector efficiency of $\eta=0.9$ in the experimental arrangement of \frf{schematic} without squeezing, i.e. a heterodyne measurement).  We observe that homodyne detection outperforms heterodyne detection as expected, although not by as much as \frf{wignerReconstruction} would indicate.  This is because an easier to reconstruct state is being probed: an amplitude $2$ (versus $3$) cat state, which has less severe Wigner oscillations.

The central theme of our paper is confirmed in \frf{fidPlot}, as we see that with sufficient squeezing the perfect efficiency homodyne limit is approached, within statistical error, for a moderate number of trials.   Detector inefficiency, finite temperature bath coupling and the intrinsic limitations of heterodyne detection are all overcome
by improving the resolution of the measured quadrature via resonant parametric amplification.
In \frf{fidPlot}, the analytical variance in the estimate of the amplified quadrature for an input coherent state can be used as an approximate proxy for the cat state reconstruction performance as measured by fidelity.  However, the reader will note that the two-step fidelity with $\chi=10$ is approximately equal to that of inefficient homodyne detection despite the quadrature estimate variance being higher.  We believe that this is due the small amount of information still present in the squeezed quadrature, which is utilized in the MaxLik algorithm.  

Here we give some further details on the simulations.  Each trial for the two-step configuration represents a simulation of the stochastic master equation (SME, see \ref{qtraj}) for finite time to produce measurement currents $dQ_X(t),dQ_Y(t)$.  The currents are then filtered and integrated similarly to \erf{filter} in order to find the POVM parameters.  The set of POVM operators representing the set of trials is then fed into the MaxLik algorithm, which iteratively moves the initial guess for the density matrix towards a more likely density matrix, as judged by the likelihood of the set of measurement currents being realized.  Once the iteration shows a change below a defined tolerance (as measured by the trace distance between the density matrices of successive iterations), the final density matrix is deemed to be the best available reconstruction of the initial state and the fidelity is calculated.  

The direct simulation of the SME is quite computationally expensive for a number of reasons: the size of the density matrix grows as the square of the used basis size, the stochastic term is of ${\cal O}(\sqrt{dt})$, the squeezing amplifies the state and, finally, the simulated measurement strength is quite strong and must be tracked carefully.  To avoid having to track the density matrix, a stochastic Schroedinger equation (SSE) is used rather than the SME.  Although the experimentalist does not have access to the specific realisation of the oscillator bath, all that is of relevance for our simulations is the production of simulated measurement currents. We can generate such currents using an SSE that simulates a measurement on the oscillator bath that will not be performed by the experimentalist.  To combat the quickly fluctuating noise, a Milstein algorithm is implemented to integrate the SSE.  To handle the amplification of the state along the anti-squeezed quadrature, an adaptively growing Fock basis is used which expands when significant probability reaches the outermost Fock states of the basis.  This is much more efficient than simulating the maximum required basis at all times.  The large measurement strength ($\mu=50$) in the final step requires the use of small time steps (about $dt=10^{-7}$).  We also use our knowledge of the time scales of the squeezing and measurement collapse from \srf{tomogram} (in particular \erf{2stepmeas2}) to determine parameters in the simulation. Firstly we select the length of time that squeezing must be carried out to achieve a large fraction of the available benefit. Secondly, most of the information in the measurement is gathered at early measurement times, so that measuring for an indefinitely long time is not necessary.  Based on these considerations, the system is squeezed and measured for $2$ units of the respective characteristic times, thus reducing the simulation duration.  Despite these techniques, the simulations would be prohibitively lengthy on a single desktop computer, so a cluster~\cite{artemis} of several hundred computing cores was used.

The purpose of simulating the SME is to obtain an integrated measurement current for each quantum trajectory.  There is an alternative method of obtaining these currents by using the POVM defined in \erf{eq:POVM}.  Given the POVM, \erf{POVMprob} then defines the probability distribution of integrated currents.  This distribution will be non-Gaussian for non-Gaussian initial states, despite the POVM itself being Gaussian.  It should then be possible to numerically sample this distribution to obtain a set of POVM operators describing the conducted trials.  This promises to be a less computationally demanding method of simulating experiments.

It is worth emphasising that the particular computational difficulty of the simulations discussed above relates to the production of artificial experimental data, not to the reconstruction procedure itself.  In an actual experiment, the measurement of the currents $I_{X,Y}$ take the place of the system simulations.  The maxLik reconstruction algorithm then needs to be performed, but its difficulty is dependent upon the basis size which is needed to estimate the unknown initial state.  This is in contrast to the simulation basis which must expand to a larger basis as the input state is amplified.  The blue-detuned system has additional intrinsic amplification so we simulated the zero-detuned system only.

\begin{figure}
	\centering
	\includegraphics[width=1\hsize]{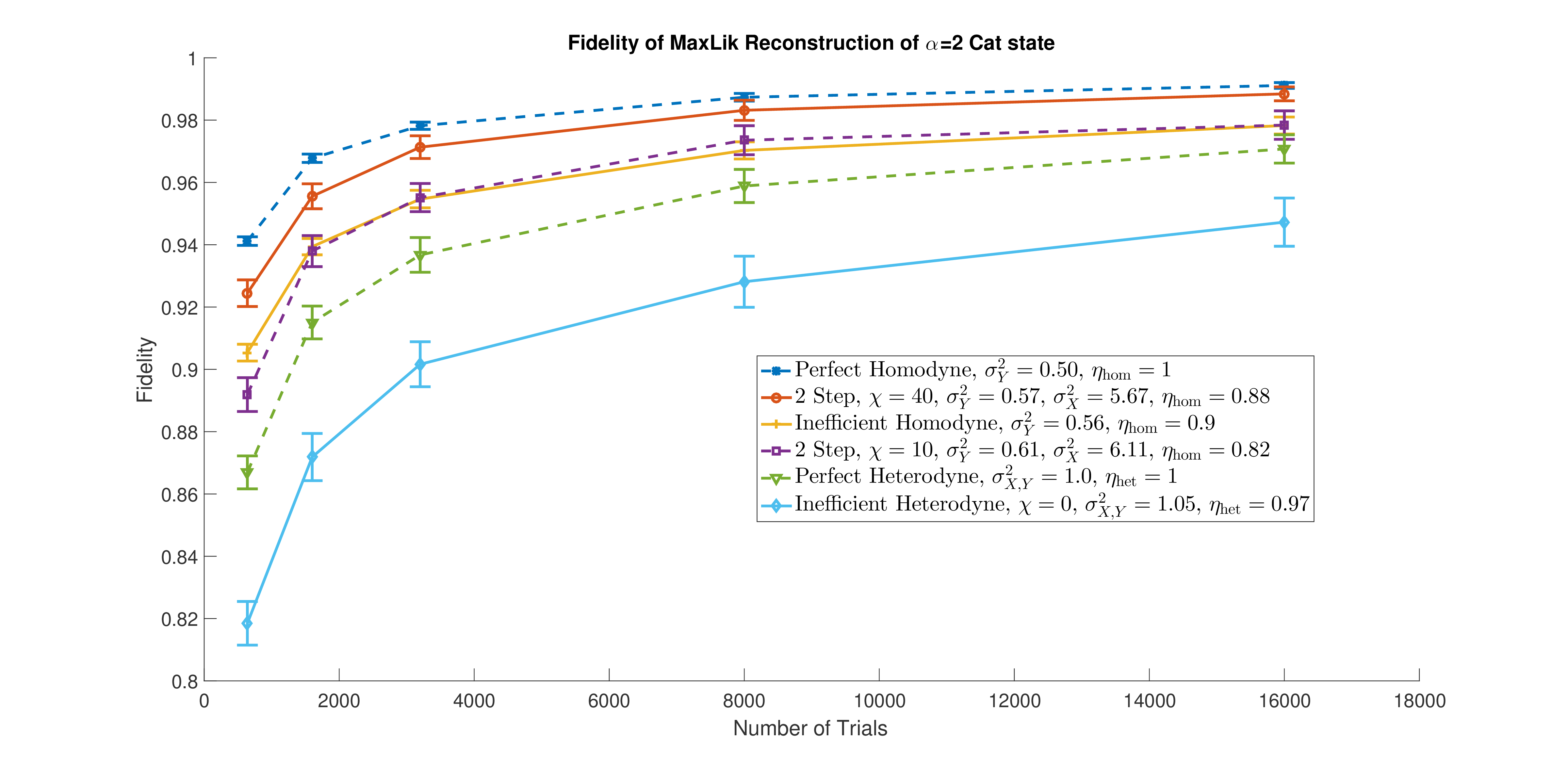}
	\caption{The average fidelity of the tomographic reconstruction of an $\alpha=2$ cat state as the number of experimental measurements is increased. The two-step protocol and zero-detuned optomechanical position measurement were used.  We also include homodyne and heterodyne measurements as benchmarks, with inefficient heterodyne detection representing no use of parametric amplification in our experiment.  The two-step protocol curves have $\gamma=0.25, \eta=0.9,n_{{\rm th}}=0.3,\mu=50, \tau=2, T=2$, where $\tau$ and $T$ are quoted in units of the characteristic time scale (see text).  Inefficient homodyne has $\eta=0.9$ while inefficient heterodyne detection has $\eta_{\rm het}=0.97$ (which corresponds to an actual detector efficiency of $\eta=0.9$ in our optomechanical arrangement of \frf{schematic}), showing the high sensitivity to imperfections for heterodyne.  50 uniformly spaced squeezing angles are used for homodyne and two-step detection. The error bars reflect statistical errors and are calculated using the standard deviation of the fidelity of the reconstructions in our simulation. In the legend, the analytical variance of the estimate of an initial coherent quadrature is given as per the formulas in \srf{tomogram}.  A conversion to an effective heterodyne or homodyne efficiency has also been provided, with the latter only applying to the anti-squeezed quadrature.
	}
	\label{fidPlot}
\end{figure}

\section{Conclusion}
\label{conc}
In this paper we have shown that parametric amplification can significantly contribute to the important goal of tomographic reconstruction of non-classical states of motion in optomechanical systems. Optomechanical measurements monitor the position of the oscillator though homodyne detection of light leaking from a coupled optical cavity. Due to the weakness of the measurement in typical experimental configurations, information is built up over many periods of motion, and thus both mechanical quadratures are obtained and the measurement is similar to optical heterodyne detection.  However, as we have shown analytically, by applying resonant parametric amplification to the oscillator, the measurement is transformed into a hybrid homodyne-heterodyne one, with the homodyne paradigm reached in the large squeezing limit.  This allows a single quadrature to be measured with variance equal to that of homodyne detection, even when there is a thermal bath and finite detector efficiency.  By placing the appropriate filter on the measurement results, the initial quadrature of the mechanical oscillator can be estimated with homodyne limited accuracy, and thus tomography performed. 

We have also shown, numerically, that homodyne, as compared to heterodyne, limited quadrature variances allow superior tomographic reconstruction of non-classical states with fast oscillating Wigner functions.  In such situations, the inclusion of parametric amplification in the experimental protocol is desirable and, as found in our simulations, leads to increased fidelity of the reconstructed state relative to the target state.  

To establish the important theoretical features that provide intuition to the analysis of all of the protocols, we began our investigation by performing (in \srf{simple})  a simplified analysis of a two-step experimental protocol (first squeezing and then measurement).  The simplification consisted of modelling the measurement step dynamics only broadly --- the finite duration measurement was replaced by an instantaneous measurement possessing some new effective detector efficiency.  Using this approach, an expression for the variance in the estimate of the anti-squeezed quadrature of a coherent state was obtained. 
As expected, it is the anti-squeezed quadrature that becomes more resolved in the presence of parametric amplification.  By inspection of the anti-squeezed quadrature estimate variance, three important thresholds of parametric amplification strength, $\chi$, were identified.  The first, and least demanding, of these, $\chi_{{\rm osc}}$, is the minimum parametric drive required to obtain a well-defined estimate of the initial state in the long squeezing-time limit. The second threshold, $\chi_{{\rm del}}$, tells us the minimum squeezing strength at which a two-step protocol is preferred, in the sense that a lower estimate variance can be achieved by including a duration of squeezing prior to optomechanical measurement. The third threshold was found to be $\chi=\chi_{{\rm het}}=2\gamma(1+n_{{\rm th}})$, which depends on the coupling to, and temperature of, the thermal bath, but is not dependent at all on the physical detector efficiency. For $\chi\gg\chi_{{\rm het}}$, the anti-squeezed quadrature estimate variance approaches $1/2$, and the composite measurement protocol (consisting of both squeezing and measurement) is as effective as an ideal homodyne measurement.  Within this same simplified framework, we also analysed misaligned parametric amplification.  This was shown to be a consideration of practical importance, with the outcome being that the squeezing should not be carried on indefinitely.  

After identification of the important parametric oscillation strength thresholds, we then performed (in \srf{tomogram}) a detailed analysis of various experimental protocols with the full measurement dynamics included.  Protocols consisting of squeezing followed by measurement (with squeezing alternatively turned off or maintained) or squeezing concurrent with measurement were analysed.  Compared with \srf{simple}, the measurement dynamics were included by giving the expressions for the effective measurement efficiency applicable to the measurement step. These expressions take into account the measurement strength, physical detector efficiency, measurement duration, bath coupling and, finally, bath temperature.  Also provided in our analyses were the filters upon the measurement results that provide the actual quadrature estimates.  

Whilst \srf{tomogram} investigated the bad-cavity zero-detuning regime, \srf{BlueDetuned} looked at what happens in the blue-detuned resolved sideband limit.  It was of interest to compare the amplification resulting from the blue-detuning to that of the parametric amplification.  We found that blue-detuning in the absence of parametric oscillation can overcome detector inefficiency but cannot improve on an ideal heterodyne measurement.  Once parametric amplification is added to the measurement protocol, it is again possible to approach the performance of an ideal homodyne measurement. The additional amplification provided by blue-detuning manifests in a lower estimate variance than zero-detuning when subjected to the same levels of squeezing, provided that the squeezing is sufficient to dominate the decohering effect of the thermal bath.

Having provided a theoretical analysis of the estimation of coherent state amplitudes in the presence of parametric amplification, we then performed (in \srf{sims}) numerical quantum state tomography for non-gaussian states.  Above the threshold $\chi_{\rm del}$, the improved resolution of the anti-squeezed quadrature with parametric amplification allows more complicated states to be reconstructed than would otherwise be possible. We showed this explicitly by simulating experimental data for the cat state $\propto \left(\ket{\alpha=3}+\ket{\alpha=-3}\right)$ with realistic optomechanical parameters. The improved reconstruction was evidenced by the Wigner functions of the reconstructions and also the fidelity with respect to the target state.

In summary, we have shown how the inclusion of parametric amplification in tomographic experimental protocols allows optomechanical measurement to be transformed from a heterodyne to homodyne type.  We have established threshold values of amplification that classify the behaviour of the variance of the estimate of an initial coherent state quadrature.  Above the largest threshold, $\chi_{{\rm het}}=2\gamma(1+n_{{\rm th}})$, the quadrature estimate variance can be made lower than that of ideal heterodyne detection.  Numerically, we then showed that this is indeed desirable for the tomography of non-classical states having fast oscillating Wigner functions.  

As creating and verifying non-classical states of motion is a major experimental goal, parametric amplification should be considered in the design of future optomechanical experiments in order to provide a greater effective homodyne efficiency.  This will lead to a reduced number of experimental trials being required.  In the circumstance of limited experimental resources, the question of how best to trade off the number of trials, the number of choices of phase of parametric drive and other parameters in the tomography experiment deserves further attention and will be the subject of future work. 

Throughout this work we have assumed that the experimental parameters are such that the dynamics of the optical cavity can be adiabatically eliminated. This choice was made because it corresponds to the regime of many experiments and because it allows for a simpler analytic treatment that helps to build intuition about the physics of optomechanical measurement enhanced by parametric amplification. We considered two parameter regimes where this was possible and found qualitiatively similar results in each case. We do not expect that our conclusions will be materially altered if this adiabatic condition does not hold. However, both our analytic and our numerical approach could be extended to the more general case and this might be interesting to investigate further.

\ack
The authors acknowledge the HPC service at The University of Sydney for providing cluster computing resources that have contributed to the research results reported within this paper.  We also acknowledge support from the Australian Research Council (ARC) via a Discovery Project (DP130103715), the Centre of Excellence in Engineered Quantum Systems (CE110001013 and CE170100009), and the University of Sydney Faculty of Science via a Postgraduate Scholarship. W.P.B. was supported by an ARC Future Fellowship (FT140100650).

\appendix
\section{Quantum Trajectories}
\label{qtraj}
In this paper, we have modelled a composite open quantum system consisting of an optical cavity coupled to a mechanical oscillator (referred to just as `oscillator'), both of which are separately in contact with the environment.  This is indicated in \frf{schematic}(a) by the cavity damping, $\kappa$, and the oscillator thermal bath, $\gamma$.  As the light exiting the cavity is correlated with the state of the oscillator, the observation of the homodyne photocurrent $I=dQ/dt$ allows its state, represented by the density matrix $\rho(t)$, to be tracked with greater accuracy than if the measurements were not made.  Note that we are considering the large $\kappa$ limit in which it is possible, both in the zero-detuned and blue-detuned regimes, to adiabatically eliminate the cavity dynamics. 
The evolution of $\rho$ is conditioned upon the values that are obtained for $dQ$ and, resultantly, $\rho$ traverses a {\it quantum trajectory} in each run of the experiment.  Quantum trajectory theory has become a ubiquitous analysis tool of monitored open quantum systems and the reader is referred to a standard text for background~\cite{WisMil10}.  The analytic core of our model is the stochastic master equation (SME) which describes how the density matrix is updated based on measurement results, $dQ$.

The Hamiltonian for the full optomechanical system in an interaction picture with respect to the cavity drive frequency and the oscillator resonance frequency is as follows
\beq
H=\omega_ma^\dagger a+\frac{\chi}{2}\sin(2\omega_mt-\theta)\left(a+ a^{\dagger }\right)^2 -\Delta b^\dagger b + g\left(a+a^\dagger\right)\left(b+b^\dagger\right),
\eeq
where $a$ is the lowering operator for the mechanical oscillator and $b$ is the lowering operator for optical cavity and we are using units where $\hbar=1$. As discussed in the main text, $\omega_m$ is the angular frequency of the mechanical oscillation, $\chi$ is the strength of the mechanical parametric drive modulating the oscillator spring constant at twice $\omega_m$, $\Delta$ is the detuning of the laser drive away from the optical cavity resonance frequency and $g$ is the optomechanical coupling. The SME describing this optomechanical interaction along with mechanical dissipation, cavity decay and homodyne detection of the output light of the cavity is as follows:
\beq
\fl d\rho = -i[H,\rho]dt +\gamma(n_{\rm th}+1)\mathcal{D}[a]\rho dt+\gamma n_{\rm th}\mathcal{D}[a^\dagger]\rho dt +
\kappa \mathcal{D}[b]\rho dt 
+\sqrt{\eta \kappa}\left(ib \rho-i\rho b^\dagger\right)dQ, 
\label{sme0}
\eeq
where the increments $dQ$ are stochastic in that they model the noisy measurement results which have a white noise background. The normalization of the measurement record is such that $dQ^2=dt$. This is the linear form of the stochastic master equation, in which the trace of $\rho$ is not preserved~\cite{GoeGra,wisQTraj,WisMil10}. In the solution to the quantum trajectory equations, $\rm{Tr}[\rho]$ evolves and, together with the statistics of the $dQ$'s, encodes the probability density of obtaining the sequence of measurement results.  The linear form of the SME has the advantage of being easier to analytically solve in order to find an explicit time evolution operator corresponding to the measurement results $dQ$. The superoperator, $\mathcal{D}$, has the action $\mathcal{D}[a]\rho=a\rho a\dg-\frac{1}{2}(a\dg a\rho+\rho a\dg a)$. Inefficient optical homodyne detection is described by $\eta$ and the phase of the local oscillator is here chosen so that information about the mechanical oscillator position is imprinted onto the measured quadrature of the output field. The coupling of the mechanics to a thermal bath (of thermal phonon occupancy $n_{{\rm th}}$) assumes a rotating wave and Markov approximation that requires $\gamma \ll \omega_m$.

As discussed in the main text, the information about the mechanical oscillation will be contained in the homodyne photocurrent $I(t)=dQ/dt$ at frequencies close to $\omega_m$. In practice, in both theoretical analysis and in experimental protocols, it is useful to demodulate this photocurrent to obtain two real demodulated photocurrents that track the quadratures of the mechanical motion. We can imagine implementing this demodulation by averaging $I(t)$ over a time $\Delta t$ as follows~\cite{WisMil10}
\begin{equation}
\Delta Q_X =\sqrt{2} \int_{t}^{t+\Delta t} \cos (\omega_m t')dQ.
\end{equation}
The normalization is chosen such that 
\begin{equation}
\langle \Delta Q_X^2\rangle = 2 \int_t^{t+\Delta t} \cos^2(\omega_m t') dt'= \Delta t ,
\end{equation}
where we have assumed that $\Delta t$ is a multiple of the oscillation period. We can regard the demodulated photocurrent as a measurement of the mechanical quadrature in a normalized white noise background, with the noise cut off at frequencies above $\omega_m$. In the adiabatically eliminated descriptions of the oscillator dynamics that follow, we will work in an interaction picture where the rate of change of $\rho$ is very slow compared to $\omega_m$ and in that case it becomes sensible to take the limit $\Delta t\rightarrow 0$ in which $\Delta Q_X$ approaches a Wiener increment $dQ_X$. We can define $dQ_Y$ in the same way as $dQ_X$ by replacing $\cos$ with $\sin$ in the above formulae. For more details of this analysis of the optomechanical measurement see Ref.~\cite{sme}.

We can simplify the description of the optomechanical measurement in two regimes of interest as discussed in the main text. The first of the these is the bad-cavity regime at zero detuning. In the limit where $\kappa\gg \omega_m,g,\gamma (n_{\rm th}+1),\chi$ and $\Delta =0$, it is straightforward to adiabatically eliminate the cavity from the SME (\ref{sme0}) to obtain an SME for the quantum state of the mechanics alone, see for example~\cite{doherty1999feedback}. The key parameter in this SME is the effective measurement rate $\mu=4g^2/\kappa$. To further simplify the master equation, we move into an interaction picture with respect to the oscillation frequency $\omega_m$ and make a new rotating wave approximation that requires $\mu, \gamma (n_{\rm th}+1),\chi \ll \omega_m$ to obtain the following master equation~\cite{PhysRevLett.107.213603,sme}:  
\bqa
\fl d\rho &=& \frac{\chi}{4}\left[e^{-i \theta}a^2-e^{i \theta}a^{\dagger 2},\rho\right]dt+[\gamma(n_{\rm th}+1)+\mu]\mathcal{D}[a]\rho dt
+(\gamma n_{\rm th}+\mu)\mathcal{D}[a^\dagger]\rho dt 
\nonumber \\
\fl&&+\sqrt{\eta\mu}\left(X \rho+\rho X\right)dQ_X\
+\sqrt{\eta\mu}\left(Y \rho +\rho Y\right)dQ_Y.
\label{sme1}
\eqa
It will frequently be convenient to note that we can absorb the parameter $\eta$ that describes the inefficient detection by defining the measurement strength $\tilde{\mu}=\eta\mu$ and the effective bath temperature, $n=n_{{\rm th}}+\mu(1-\eta)/\gamma$. In terms of these parameters we have 
\begin{eqnarray}
\fl d\rho &=& \frac{\chi}{4}\left[e^{-i \theta}a^2-e^{i \theta}a^{\dagger 2},\rho\right]dt+[\gamma(n+1)+\tilde{\mu}]\mathcal{D}[a]\rho dt+(\gamma n+\tilde{\mu})\mathcal{D}[a^\dagger]\rho dt \nonumber \\
\fl &&+\sqrt{\tilde{\mu}}\left(X \rho+\rho X\right)dQ_X+\sqrt{\tilde{\mu}}\left(Y \rho +\rho Y\right)dQ_Y.
\label{sme}
\end{eqnarray}

In order to determine the optimal filter for the measurement current, and also analyse the tomographic performance of various experimental procedures, the POVM element describing the composite measurement up to the time $t$ is required.  This operator, which allows the statistics of the measurement to be found, follows straightforwardly from the analytic solution of \erf{sme}.  Recently, two of the authors have given a general method of solution for Gaussian bosonic linear stochastic master equations such as this~\cite{warDoh}. The solution is a generalisation of methods that have been used to find solutions to linear stochastic Schroedinger equations in the past~\cite{jacLin,wisQTraj,herkommer1996localization}.

If $\rho_0$ is the initial state of the system, we find a solution having the general form
\begin{equation}
\rho(t)=e^{Z} \mathcal{E}_{X_{\rm est},Y_{\rm est}}[\rho_0], \label{eq:smesoln}
\end{equation}
where $\mathcal{E}_{X_{\rm est},Y_{\rm est}}$ is a Gaussian completely positive map that depends on only two real-valued functions of the measurement record, $X_{\rm est}$ and $Y_{\rm est}$ as defined in equation (\ref{filter2}). $Z$ is a functional of the measurement currents $I_X$ and $I_Y$ that determines a numerical prefactor in (\ref{eq:smesoln}). We are interested in the statistics of the measurement outcomes $X_{\rm est}$ and $Y_{\rm est}$ and can neglect $Z$ since it provides no information (that is not contained in $X_{\rm est}$ and $Y_{\rm est}$) about the initial state. The probability density for these initial state estimates is proportional to ${\rm Tr}[\rho(t)]$. Specifically, the general theory~\cite{wisQTraj} implies that
\begin{equation}
P(X_{\rm est},Y_{\rm est}|\rho_0) = {\rm Tr}[\mathcal{E}_{X_{\rm est},Y_{\rm est}}[\rho_0] ]\int e^Z P_{\rm ost}(X_{\rm est},Y_{\rm est},Z) dZ
\end{equation}
where the \emph{ostensible probability distribution} $P_{\rm ost}(X_{\rm est},Y_{\rm est},Z) $ is the joint probability distribution that $X_{\rm est},Y_{\rm est},$ and $Z$ would have if the measurement records $I_X(t)$ and $I_Y(t)$ were both mean zero white noise processes. $Z$ is a non-Gaussian random variable in the ostensible distribution and is correlated with $X_{\rm est}$ and $Y_{\rm est}$, so the integral here may appear complicated. However it can be shown that $Z$ is the sum of squares of random variables that have Gaussian distributions under the ostensible distribution and consequently the above integral can be evaluated in the usual way for Gaussian integrals and evaluates to a Gaussian function of $X_{\rm est}$ and $Y_{\rm est}$. 

The probability density $P(X_{\rm est},Y_{\rm est}|\rho_0)$ determines the POVM for the composite measurement $W_I$ through the formula 
\beq
P(X_{\rm est},Y_{\rm est}|\rho_0) = {\rm Tr}[W_I\rho_0].
\label{POVMprob}
\eeq
In the main text we specify the POVM by considering the case where $\rho_0=|\alpha_0\rangle\langle \alpha_0|$ so that $P(X_{\rm est},Y_{\rm est}|\rho_0)$ is a Gaussian distribution (\ref{eq:POVM}) for $X_{\rm est}$ and $Y_{\rm est}$. As mentioned in the main text, this uniquely specifies the POVM just as the Q-function uniquely specifies a density matrix. The method of obtaining expressions for the parameters of this Gaussian in terms of the system parameters is detailed in~\cite{warDoh}, to which we refer the interested reader.

In the second regime of interest, being the blue-detuned resolved-sideband regime in which $\Delta=\omega_m$ and $\omega_m\gg \kappa \gg g,\chi$, we can go into an interaction picture with respect to $\omega_m$ and $\Delta$ prior to the adiabatic elimination, and make a rotating wave approximation that removes the fast rotating terms in the Hamiltonian. The resulting optomechanical coupling is of the form $g(ab+a^\dagger b^\dagger)$.  Making the adiabatic approximation as before, we now find the modified linear SME
\bqa
\fl d\rho &=&\frac{\chi}{4}\left[e^{-i \theta}a^2-e^{i \theta}a^{\dagger 2},\rho\right]dt+\gamma(n_{{\rm th}}+1)\mathcal{D}[a]\rho dt+(\gamma n_{{\rm th}}+\mu)\mathcal{D}[a^\dagger]\rho dt\nonumber\\
\fl &&+\sqrt{\eta\mu}\left(dZ a^\dag \rho+dZ^{*}\rho a\right),
\label{BDsme}
\eqa
where we have introduced the complex combination of measurement increments $dZ=\left(dQ_X+idQ_Y\right)/\sqrt{2}$ for convenience.  The reader will note several differences to the zero-detuned SME.  The measurement back action appears only in the dissipation term for $a^\dag$. The details of the measurement update terms proportional to $dQ_X$ and $dQ_Y$ are also different. They have the same form as the usual heterodyne detection stochastic master equation~\cite{WisMil10} but with $a$ replaced by $a^\dagger$. This distinction relates to the fact that the position measurement signal from the mechanics is amplified by the optomechanical coupling, improving the signal to noise with which the initial mechanical state can be inferred from the results of the optical homodyne measurement. As is the case for heterodyne detection, the two demodulated currents $I_X$ and $I_Y$ carry information about the corresponding mechanical quadratures. Also the detector efficiency, $\eta$, has been explicitly included as it cannot be absorbed into the other parameters.  The POVM for blue-detuning is found using the same methods as for zero-detuning and is also of the form \erf{eq:POVM}.

\section*{References}
\bibliographystyle{iopart-num}
\bibliography{bibliographyNJPandrew}

\end{document}